\documentclass[twoside,twocolumn,9pt]{article}
\usepackage[utf8]{inputenc}
\usepackage{extsizes}
\usepackage[super,sort&compress,comma]{natbib} 
\usepackage[version=3]{mhchem}
\usepackage[left=1.5cm, right=1.5cm, top=1.785cm, bottom=2.0cm]{geometry}
\usepackage{balance}
\usepackage{times,mathptmx}
\usepackage{sectsty}
\usepackage{graphicx} 
\usepackage{lastpage}
\usepackage[format=plain,justification=justified,singlelinecheck=false,font={stretch=1.125,small,sf},labelfont=bf,labelsep=space]{caption}
\usepackage{float}
\usepackage{fancyhdr}
\usepackage{fnpos}
\usepackage[english]{babel}
\addto{\captionsenglish}{%
  
}
\usepackage{array}
\usepackage{droidsans}
\usepackage{charter}
\usepackage[T1]{fontenc}
\usepackage[usenames,dvipsnames]{xcolor}
\usepackage{setspace}
\usepackage[compact]{titlesec}
\usepackage{hyperref}

\usepackage{epstopdf}

\definecolor{cream}{RGB}{222,217,201}

\begin{document}

\pagestyle{fancy}
\thispagestyle{plain}
\fancypagestyle{plain}{

\fancyhead[C]{\includegraphics[width=18.5cm]{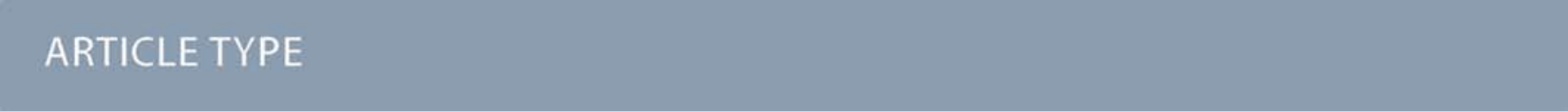}}
\fancyhead[L]{\hspace{0cm}\vspace{1.5cm}\includegraphics[height=30pt]{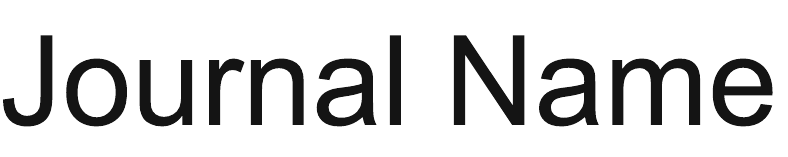}}
\fancyhead[R]{\hspace{0cm}\vspace{1.7cm}\includegraphics[height=55pt]{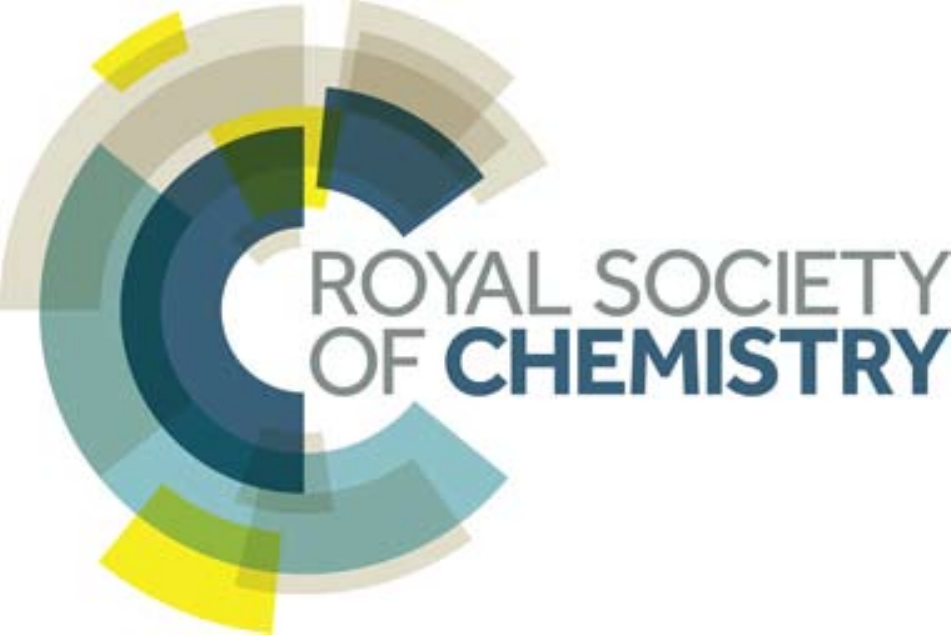}}
\renewcommand{\headrulewidth}{0pt}
}

\makeFNbottom
\makeatletter
\renewcommand\LARGE{\@setfontsize\LARGE{15pt}{17}}
\renewcommand\Large{\@setfontsize\Large{12pt}{14}}
\renewcommand\large{\@setfontsize\large{10pt}{12}}
\renewcommand\footnotesize{\@setfontsize\footnotesize{7pt}{10}}
\makeatother

\renewcommand{\thefootnote}{\fnsymbol{footnote}}
\renewcommand\footnoterule{\vspace*{1pt}%
\color{cream}\hrule width 3.5in height 0.4pt \color{black}\vspace*{5pt}} 
\setcounter{secnumdepth}{5}

\makeatletter 
\renewcommand\@biblabel[1]{#1}            
\renewcommand\@makefntext[1]%
{\noindent\makebox[0pt][r]{\@thefnmark\,}#1}
\makeatother 
\renewcommand{\figurename}{\small{Fig.}~}
\sectionfont{\sffamily\Large}
\subsectionfont{\normalsize}
\subsubsectionfont{\bf}
\setstretch{1.125} 
\setlength{\skip\footins}{0.8cm}
\setlength{\footnotesep}{0.25cm}
\setlength{\jot}{10pt}
\titlespacing*{\section}{0pt}{4pt}{4pt}
\titlespacing*{\subsection}{0pt}{15pt}{1pt}

\fancyfoot{}
\fancyfoot[LO,RE]{\vspace{-7.1pt}\includegraphics[height=9pt]{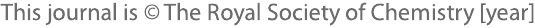}}
\fancyfoot[CO]{\vspace{-7.1pt}\hspace{13.2cm}\includegraphics{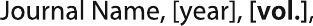}}
\fancyfoot[CE]{\vspace{-7.2pt}\hspace{-14.2cm}\includegraphics{head_foot/RF}}
\fancyfoot[RO]{\footnotesize{\sffamily{1--\pageref{LastPage} ~\textbar  \hspace{2pt}\thepage}}}
\fancyfoot[LE]{\footnotesize{\sffamily{\thepage~\textbar\hspace{3.45cm} 1--\pageref{LastPage}}}}
\fancyhead{}
\renewcommand{\headrulewidth}{0pt} 
\renewcommand{\footrulewidth}{0pt}
\setlength{\arrayrulewidth}{1pt}
\setlength{\columnsep}{6.5mm}
\setlength\bibsep{1pt}

\makeatletter 
\newlength{\figrulesep} 
\setlength{\figrulesep}{0.5\textfloatsep} 

\newcommand{\topfigrule}{\vspace*{-1pt}%
\noindent{\color{cream}\rule[-\figrulesep]{\columnwidth}{1.5pt}} }

\newcommand{\botfigrule}{\vspace*{-2pt}%
\noindent{\color{cream}\rule[\figrulesep]{\columnwidth}{1.5pt}} }

\newcommand{\dblfigrule}{\vspace*{-1pt}%
\noindent{\color{cream}\rule[-\figrulesep]{\textwidth}{1.5pt}} }

\makeatother

\twocolumn[
  \begin{@twocolumnfalse}
\vspace{3cm}
\sffamily
\begin{tabular}{m{4.5cm} p{13.5cm} }

\includegraphics{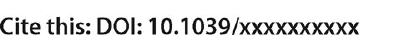} & \noindent\LARGE{\textbf{Curvature dynamics and long-range effects on fluid-fluid interfaces with colloids}} \\
\vspace{0.3cm} & \vspace{0.3cm} \\

 & \noindent\large{A. Tiribocchi,$^{\ast}$\textit{$^{a,b}$} F. Bonaccorso,\textit{$^{a}$} M. Lauricella,\textit{$^{b}$}, S. Melchionna\textit{$^{c}$}, A. Montessori\textit{$^{d}$}, S. Succi\textit{$^{a,b,e}$}} \\

\includegraphics{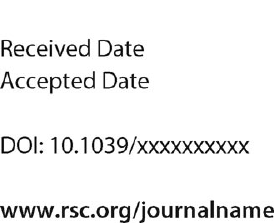} & \noindent\normalsize{We investigate the  dynamics of a phase-separating binary fluid, containing colloidal dumbbells anchored to the fluid-fluid interface. Extensive Lattice Boltzmann-Immersed Boundary method simulations reveal that the presence of soft dumbbells can significantly affect the curvature dynamics of the interface between phase-separating fluids, even though the coarsening dynamics is left nearly unchanged. In addition, our results show that the  curvature dynamics exhibits distinct non-local effects, which might be exploited for the design of new soft mesoscale materials. We point out that the inspection of the statistical dynamics of the curvature can disclose new insights into local inhomogeneities of the binary fluid configuration, as a function of the volume fraction and aspect ratio of the dumbbells.
}

\end{tabular}

 \end{@twocolumnfalse} \vspace{0.6cm}

  ]

\renewcommand*\rmdefault{bch}\normalfont\upshape
\rmfamily
\section*{}
\vspace{-1cm}


\footnotetext{\textit{$^{a}$ Center for Life Nano Science$@$La Sapienza, Istituto Italiano di Tecnologia, 00161 Roma, Italy; E-mail: adriano.tiribocchi@iit.it}}
\footnotetext{\textit{$^{b}$ Istituto per le Applicazioni del Calcolo CNR, via dei Taurini 19, 00185, Rome, Italy; E-mail: m.lauricella@iac.cnr.it}}
\footnotetext{\textit{$^{c}$ ISC-CNR, Istituto Sistemi Complessi, Università Sapienza, P.le A. Moro 2, 00185 Rome, Italy; E-mail: simone.melchionna@isc.cnr.it}}
\footnotetext{\textit{$^{d}$ Department of Engineering, University of Rome, ``Roma Tre'' Via Vito Volterra 62, 00146 Rome, Italy; E-mail: andrea.montessori@uniroma3.it}}
\footnotetext{\textit{$^{e}$ Institute for Applied Computational Science, John A. Paulson School of Engineering and Applied Sciences, Harvard University, Cambridge, Massachusetts 02138, USA; E-mail: s.succi@iac.cnr.it}}






\section{Introduction}

In recent years major research effort has been directed to the design of novel mesoscale soft materials. Of particular relevance is the case of soft-glassy materials (which include systems like emulsions, foams and gels), which have found applications in several sectors of modern industry, such as chemical and food processing, manufacturing and biomedical, to name but a few \cite{Fern,Mezz,Piaz}. Besides their technological importance, these systems hold an enormous theoretical interest, due to their ability to support intriguing non-equilibrium effects, such as long-time relaxation, yield-stress behavior and highly non-Newtonian dynamics \cite{Berhier,Cohen,Sollich,Cates2}. These features play a crucial role in designing and shaping up novel soft porous materials, with enhanced mechanical properties, thereby opening exciting scenarios in the realisation of new states of matter.   

Importantly, the stability and the morphology of these metastable structures are crucially affected by the dynamics of complex interfaces, which interact with the surrounding material in a highly non-trivial fashion. A remarkable example in point is provided by the bijel \cite{Cates2,Herzig,Huang}, an amorphous soft-solid material, in which a densely packed monolayer of colloidal particles is sequestered at the fluid-fluid interface of a bicontinuous fluid domain. Parameters like particle size, interface width and wetting conditions, are key to control the thermodynamic process history, to improve mechanical stability and to understand the rheological behavior \cite{Cates2,Herzig,Harting,Harting2}. Pickering emulsions \cite{Picker}, i.e. droplets stabilized by colloidal particles sitting at the fluid-fluid interface, represent a further example in which the interplay between the local mesoscopic structure of the interface and the particle size crucially affect the material design \cite{Cheval,Melle,Harting,Harting2}. 

In spite of the tremendous progress in the design of new soft materials, the exact dynamics governing the interface motion is still only partially understood, even for relatively simple systems, such as a binary fluid mixture \cite{Bray}. A further complication emerges when solid objects, such as spherical colloids, are included \cite{Jaya,Stratford}. The basic question arising in this framework is: how does the fluid-fluid interface dynamics affect the mechanical properties of such materials? And, more specifically, how does its curvature impact on the structure and on the organization of the adsorbed particles?

In this work, we address this question by means of large-scale Lattice Boltzmann simulations \cite{Succi2,Succi3,Chen,Succi5,Bernaschi}, to investigate the fluid-fluid interface dynamics of a binary fluid mixture containing colloidal particles. These colloidal particles are modelled as dumbbells confined at the fluid-fluid interface and simulated by means of an immersed boundary method  \cite{Melchionna}. Unlike other anisotropic particles previously considered\cite{Gunther,Gunther2,Clegg}, the present ones can mimick more effectively the physics of an amphiphilic ``super-surfactant'' (as, in our model, their size is either comparable or larger than that of the fluid interface, usually of the order of $1\mu$m), whose constituents (resembling the molecules) are anisotropic shaped.

By varying their volume fraction and aspect ratio, we find that they have negligible effects on the interface shrinking (coarsening process), even at relatively high volume fraction, but significantly affect the dynamics of the interface curvature. A careful inspection of the time evolution of the probability density functions (pdfs) of the curvature, unveils a spontaneous transition from an initial broadly-shaped distribution to a highly localised, steady-state one, whose average and variance depend significantly on particle volume fraction and aspect ratio. The same trend is also captured by the relative entropy of pdf, a quantity providing a global measure of the ``distance'' of  steady-state configuration from the initial one. 

Our results also show that the curvature dynamics displays a strongly super-diffusive behavior (even in the absence of colloids), thus suggesting that the fluid-fluid interface acts as a long-range correlator within the binary fluid, meaning by this that a perturbation at a given location on the interface propagates faster along the interface than in the bulk fluid, thereby providing  a fast-track non-local communication mechanism. This could be probably tested in a lab experiment by comparing the speed of such perturbation, generated by applying a force at the fluid interface, with the one measured in the bulk fluid. Inspection of the dynamics of the pdf's of the curvature suggests that such non-local dynamics can be arguably described by a fractional Fokker-Planck equation. These results point to the interface curvature as to a valuable observable, potentially capable of capturing new aspects and mechanisms of the interface dynamics, hence offering new clues for the design of novel soft mesoscale materials. For instance, one may envisage to tailor, in a programmable manner, shape or specific mechanical properties of portions of materials placed far apart from a local source of stimulation applied at the fluid interface (by carefully tuning, for example, its stiffness or the fluid viscosity). This would likely provide a viable strategy to modify the design of a soft material {\it en route} and on demand, by simultaneously minimizing manufacturing process and defects on the final product, with a potential reduction of costs of material fabrication. Self-sustaining composite soft structures free from scaffold support, produced by means of 3d printers \cite{Colosi}, may be the suitable system to investigate such effects.

The paper is organised as follows. In Section II we describe the numerical model, with details on the implementation of the interface-confined colloids and the corresponding equations of motion. In Section III  we first investigate, as benchmark tests, the effect that a single dumbbell produces on a flat fluid interface and how a medium/low dumbbell volume fraction modifies the curvature of an isolated fluid droplet. Afterwards we provide a quantitative description of the fluid-fluid interface dynamics observed during phase separation, with and without the presence of colloids. Later on, we report the results on the dynamics of the interface curvature, both in terms of  the first and second order moments (average and variance, respectively) and of the full probability distribution function. A discussion on the dependence of the steady-state pdfs on the volume fraction and dumbbell aspect ratio is also provided. Finally, we present some concluding remarks.

\section{Model and equations}

Here we describe the physics and the modeling of a phase separating binary fluid mixture, with a collection of colloidal dumbbells confined at its interface. The dynamics of the binary fluid is governed by the continuity and the Navier-Stokes equations, numerically solved by means of Lattice Boltzmann (LB) simulations \cite{Succi2,Succi3,Chen}. Such method has proven capable to simulate the physics of multiphase flows (such as surface tension \cite{Succi3} and disjoining pressure \cite{Korner,Sbragaglia,Sukop}),  as well as of other assorted complex fluids, such as porous \cite{Succi4} and soft glassy materials \cite{Benzi}, polymers \cite{Duenweg}, liquid crystals \cite{Henrich,Foffano,Marenduzzo}, and even blood \cite{Melchionna,Pontrelli}. We initially summarise the LB approach adopted to study large-scale blood flows \cite{Melchionna} and then we discuss the inclusion of  colloidal dumbbells. 

\subsection{Lattice Boltzmann model of the binary fluid}

LB is built starting from a discrete set of distribution functions $f_p({\bf x}, t)$, each of which represents the probability to find a fluid particle $p$ at time $t$ on a lattice site ${\bf x}$ and travelling with discrete velocity ${\bf c}_p$. The dynamics of $f_p$ over a timestep $\Delta t$ is governed by a discrete Boltzmann equation of the form:   
\begin{equation}\label{streaming}
f_p({\bf x}+\Delta t{\bf c}_p, t + \Delta t)= f^*_p({\bf x},t),
\end{equation}
where the right-hand side describes the effect of streaming, namely the motion of free particles along straight trajectories, whereas $f_p^*({\bf x},t)$ represents the post-collisional population, given by
\begin{equation}\label{collision}
f_p^*=f_p-\frac{\Delta t}{\tau}(f_p-f_p^{eq})+\Delta t\Delta f_p^{drag}.
\end{equation}
Here, $\tau$ is a characteristic time, setting the typical relaxation timescale of $f_p$ towards its local equilibrium \cite{Succi2}, and $f_p^{eq}$ are local equilibrium distribution functions written as a second-order expansion in the fluid velocity ${\bf u}$,
\begin{equation}
  f_p^{eq}=w_p\rho\left[1+\frac{{\bf u}\cdot{\bf c}_p}{c_s^2}+\frac{{\bf uu}:({\bf c}_p{\bf c}_p-c_s^2{\bf I})}{2c_s^4}\right],
\end{equation}
where $c_s=1/{\sqrt 3}$ is the speed of sound, $\rho$ is the fluid density, ${\bf I}$ is the unit matrix and $w_p$ is a normalised set of weights. In this work, we employ the $D3Q19$ lattice scheme, namely a three dimensional lattice with $p=0,...,18$, with $w_p=1/3$ for ${\bf c}_0=(0,0,0)$, $w_p = 1/18$ for ${\bf c}_{1,...,6}=(\pm 1, 0, 0), (0,\pm 1, 0), (0,0,\pm 1)$ (i.e. vectors connecting nearest lattice neighbors),  and $w_p = 1/36$ for ${\bf c}_{7,...,18} =(\pm 1,\pm 1, 0), (\pm 1, 0, \pm 1), (0, \pm 1, \pm 1)$ (i.e. vectors linking next-nearest lattice neighbors). The last term of Eq.(\ref{collision}), $\Delta f_p^{drag}$, can be expressed as a second order Taylor expansion in the lattice velocity \cite{Guo}
\begin{equation}
\Delta f_p^{drag}=w_p\rho\left[\frac{{\bf G}\cdot{\bf c}_p}{c_s^2}+\frac{({\bf G}\cdot{\bf c}_p)({\bf u}\cdot{\bf c}_p)-c_s^2{\bf G}\cdot{\bf u}}{2c_s^4}\right]\Delta t,  
\end{equation}
where ${\bf G}={\bf G}_1+{\bf G}_2$ is the sum of an interaction force ${\bf G}_1$ between fluid components $\lambda$ and $\bar{\lambda}$ and of a local coupling force ${\bf G}_2$ between suspended particles and fluid (see next section). By following the Shan-Chen approach \cite{Shan,Shan2} the former term is given by
\begin{equation}
{\bf G}_1({\bf x})=-g_{\lambda\bar{\lambda}}\psi^{\lambda}({\bf x})\sum_pw_p\psi^{\lambda}({\bf x}+{\bf c}_p){\bf c}_p,
\end{equation}
where the coupling constant $g_{\lambda\bar{\lambda}}$ controls the interaction strength between fluid particles. A positive (negative) value codes for repulsive (attarctive) forces, respectively. To promote phase separation, in our simulation we set a positive value for $g_{\lambda\bar{\lambda}}$. In addition, this term, together with the relaxation time $\tau$, controls the surface tension, and is vaguely similar to the elastic constant appearing in free-energy lattice Boltzmann models \cite{Orlandini,Tiribocchi}. Note that if $\lambda=\bar{\lambda}$ (namely a one-component fluid), $g_{\lambda\bar{\lambda}}=0$. Finally, $\psi$ is an effective density term which in our case coincides with the physical density of the fluid,  $\psi(\rho)=\rho$. Such simple functional form  ensures mechanical stability and thermodynamic consistency of the system \cite{Chen}. The second term ${\bf G}_2$ will be discussed in the next section.

The local fluid density $\rho$, the fluid-momentum $\rho{\bf u}$ and the pressure tensor ${\bf P}$, can be computed directly from the kinetic moments of the populations $f_p$ by linear and local summations:
\begin{eqnarray}
\rho&=&\sum_pf_p\\
\rho{\bf u}&=&\sum_pf_p{\bf c}_p+\frac{\Delta t}{2}\rho{\bf G}\\
{\bf P}&=&\sum_pf_p{\bf c}_p{\bf c}_p.
\end{eqnarray}
The fluid pressure is given by the diagonal terms of the pressure tensor ${\bf P}$, while the stress $\Pi$ stems from the off-diagonal terms, and is related to the distribution function by
\begin{equation}
\Pi\equiv\nu\rho(\partial{\bf u}+\partial{\bf u}^T)=\frac{\nu}{c_s^2\tau}\sum_p{\bf c}_p{\bf c}_p(f_p-f_p^{eq}).
\end{equation}
By means of a Chapman-Enskog expansion of the distribution function in its time and space derivatives \cite{Chen}, the continuity and the Navier-Stokes equations in the incompressible limit are finally recovered
\begin{eqnarray}
\nabla\cdot {\bf u}&=&0,\\
\frac{\partial{\bf u}}{\partial t}+({\bf u}\cdot\nabla){\bf u}&=&-\frac{1}{\rho}\nabla P + \nu\nabla^2{\bf u} +{\bf F},
\end{eqnarray}
with $P$ isotropic pressure, $\nu=c^2_s(\tau-\Delta t/2)$ kinematic viscosity, and ${\bf F}$ body force corresponding to $\Delta f_p^{drag}$. Note that thermal fluctuations are not included in the model.

\subsection{Colloidal dumbbells}
\begin{figure}
\centerline{\includegraphics[width=0.5\textwidth]{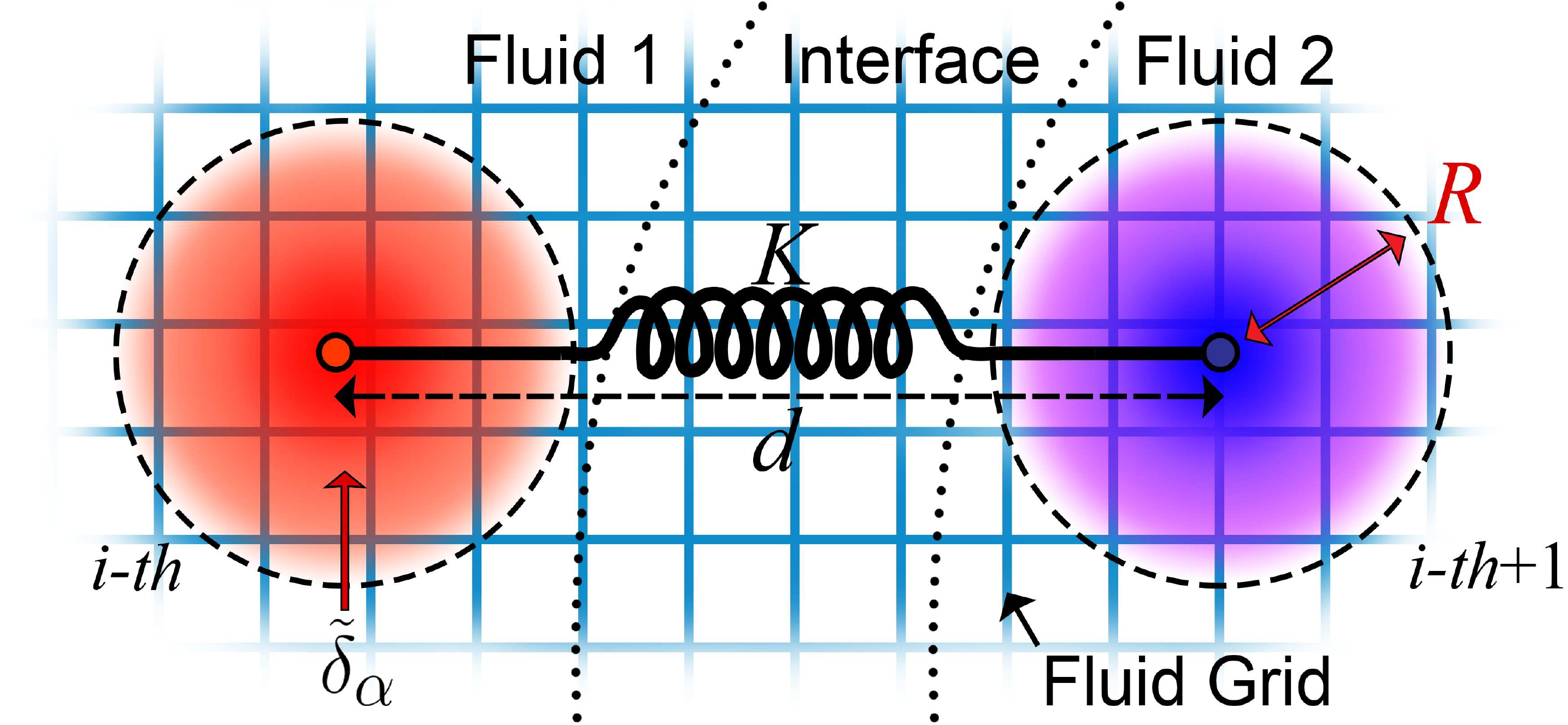}}
\caption{A cartoon of the dumbbell in our model. Two spherical beads of radius $R$, placed at equilibrium distance $d$, interact through an elastic potential, represented by a spring of elastic constant $K$. Beads are modeled by means of an immersed boundary method, which employes an appropriate interpolating function $\tilde{\delta}_{\alpha}$ to generate a spherically symmetric shaped particle (see text). Dotted lines represent the fluid-fluid interface where dumbbells are anchored. On the background, the grid where populations $f_p$ are defined.}
\label{fig1A}
\end{figure}

By following Ref. \cite{Melchionna}, we model our noise-free colloidal dumbbells by means of two spherical beads of radius $R$, placed at equilibrium distance $d$ between their centers of mass, and interacting through an elastic potential with elastic constant $K$. These parameters control the aspect ratio of the dumbbell, defined as $A=(d+2R)/2R$, a key quantity in our simulations. In Fig.\ref{fig1A} we show a sketch of the dumbbell.

The bead of the dumbbell is modeled via an immersed boundary method \cite{Melchionna}, in which its positions and velocities are indicated respectively by ${\bf R}_i$ and ${\bf V}_i$, with $i=1,...,N_b$ and $N_b$ being the number of beads. The spherical shape of each bead can be described via the following interpolating (or shape) function:
\begin{equation}
\tilde{\delta}({\bf x})={\displaystyle\prod_{\alpha=x,y,z}}\tilde{\delta}_{\alpha}[({\bf x})_{\alpha}],
\end{equation}
with
$$
\tilde{\delta}_{\alpha}(x_{\alpha})\equiv
\begin{cases}
  \frac{1}{8}\left(5-4|x_{\alpha}|/\xi-\sqrt{1+8|x_{\alpha}|/\xi-16x^2_{\alpha}/\xi^2}\right)\\ \hspace{3cm}\mbox{if $|x_{\alpha}|/\xi\leq 0.5$}\\
  \frac{1}{8}\left(3-4|x_{\alpha}|/\xi-\sqrt{-7+24|x_{\alpha}|/\xi-16x^2_{\alpha}/\xi^2}\right)\\ \hspace{3cm}\mbox{if $0.5<|x_{\alpha}|/\xi\leq 1$}\\
  0 \hspace{2.83cm}\mbox{if $|x_{\alpha}|/\xi>1$},
\end{cases}
$$
where $\alpha=x,y,z$ indicates the Cartesian components of the underlying Eulerian mesh. For $\xi=2$ the shape function generates a spherically symmetric diffused particle, extending over $4^3=64$ mesh points. The function is also normalized when summed over Cartesian mesh points ${\bf x}$, $\sum_x\tilde{\delta}({\bf x}-{\bf X})=1$ (for any continuous displacement ${\bf X}$), and obeys to $\sum_x(x_{\alpha}-X_{\alpha})\partial_{\beta}\tilde{\delta}({\bf x}-{\bf X})=-\delta_{\alpha\beta}$ \cite{Melchionna}.

In this study, we consider only the translational coupling between the dumbbell and the surrounding fluid and neglect the rotational one. This approximation would not generally inhibit an out-of-plane dumbbell rotation (triggered, for instance, by the fluid flow), but it does not allow an explicit control of the body rotational response (i.e. of the torque exerted by the fluid), which has been found to generate capillary effects at the fluid interface when anisotropic particles are adsorbed\cite{Davies}. In spite of this simplification, such minimal description is sufficient to capture the relevant physics of the curvature dynamics of the fluid interface. The functional form of the translational coupling term is given by:
\begin{equation}
\phi({\bf x},i)=-\gamma_T\tilde{\delta}({\bf x}-{\bf R}_i)[{\bf V}_i-{\bf u}({\bf x})]=-\gamma_T\tilde{\delta}_i({\bf V}_i-{\bf u}),  
\end{equation}
where $\gamma_T$ is a translational coupling coefficient and  $\tilde{\delta}_i\equiv\tilde{\delta}({\bf x}-{\bf R}_i)$. Hence, the hydrodynamic force acting on each bead can be derived by integrating (actually a discrete sum over the lattice) over the particle volumetric extension
\begin{equation}
{\bf F}_i=\sum_x\phi({\bf x},i)=-\gamma_T({\bf V}-\tilde{\bf u}_i),
\end{equation}
where
\begin{equation}
\tilde{\bf u}_i=\sum_x\tilde{\delta}{\bf u}.
\end{equation}
The action of the forces ${\bf F}_i$ is counterbalanced by the opposite reaction on the fluid side, which is given by
\begin{equation}
{\bf G}_2({\bf x})=-\sum_i{\bf F}_i\tilde{\delta}_i.
\end{equation}

Besides the coupling between the dumbbell and the fluid, a further ingredient is needed to correctly model the dynamics of the system, namely the dumbbell-dumbbell interaction. This is described by means of a Weeks-Chandler-Andersen (WCA) potential \cite{Weeks},  which inhibits interpenetration of pairs of dumbbells by introducing a repulsive interaction. Its functional form stems from the Lennard-Jones potential and is given by
$$
V_{WCA}(r)=\begin{cases}
4\epsilon[(\frac{\sigma}{r})^{12}-(\frac{\sigma}{r})^6]+\epsilon \hspace{0.2cm}\mbox{if} \hspace{0.2cm} r < 2^{1/6}\sigma\\
0 \hspace{2.8cm} \mbox{if} \hspace{0.2cm} r \geq 2^{1/6}\sigma
\end{cases}
$$
where $\epsilon$ sets the energy scale (the well depth), $r$ is the center-to-center separation between two particles, and $\sigma$ is the value of $r$ at which $V_{WCA}(r)=0$. This is, to a good approximation, the diameter of the bead.

\subsection{Solvation force}
In order to anchor dumbbells at the fluid-fluid interface, we add a solvation force, coupling each dumbbell to the phase of the fluid, to the forcing term ${\bf G}$. This term consists of two contributions, the first representing the force exerted by each bead of the dumbbell on the surrounding fluid, given by
\begin{equation}
{\bf F}_{i}^b=\sum_{k}\sum_{\alpha}\Lambda_k\rho^k_{\alpha}\nabla{\tilde\delta}_{i}({\bf x}).
\end{equation}
Here $k=1,2$ indicates the two phases of the binary fluid ($\lambda$ and $\bar{\lambda}$) and $\alpha$ runs over the lattice sites located inside the spherical bead. The parameter $\Lambda_k$ is a constant whose absolute value gauges the interaction strength of each bead with each component of the fluid, while its sign determines whether the interaction is either attractive or repulsive. If, for instance, the bead is surrounded by the fluid $\lambda$, $\Lambda_k$ is positive (i.e. attractive) for this one, and negative (i.e. repulsive) for $\bar{\lambda}$. Finally, the second contribution to the total force accounts for the reaction of the surrounding fluid, and is given by
\begin{equation}
{\bf G}_{3}({\bf x})=-\sum_i\sum_{k}\sum_{\alpha}\Lambda_k\rho^k_{\alpha}\nabla{\tilde\delta}_{i}({\bf x}).
\end{equation}

\subsection{Numerical aspects}

Before discussing our results, we first report the numerical details of our study. Simulations are performed on a periodic cubic lattice of linear size $L=256$, with the following parameters: $\Delta x=1$, $\Delta t=1$, $\tau=1$, $g_{\lambda\bar{\lambda}}=1.2$, $\gamma_T=0.1$. The values of $\tau$ and $g_{\lambda\bar{\lambda}}$ set an interface width varying from $5-7$ lattice sites. We also kept fixed the energy scale of the WCA potential $\epsilon=10^{-4}$, the coupling constant of the solvation force $\Lambda_k=\pm 0.1$, the bead radius $R=2$ and the elastic constant $K \simeq 1$, and varied the center of mass distance $d$, in order to change the aspect ratio $A$. The particle volume fraction is then defined as $V_f=\frac{\frac{4}{3}\pi R^3 2N_b}{L^3}$, and, for a fixed value of $A$, is varied by changing the number of beads $N_b$.

The phase separation is simulated starting from an initial configuration in which the two components $\lambda$ and $\bar{\lambda}$ of the binary fluid are mixed (at temperature $T$ over the critical value $T_c$ of the coexistence region of the phase diagram~\cite{Bray}) and particles of predefined aspect ratio $A$ are randomly distributed in the lattice. Subsequently, the mixture is quenched down to a temperature $T<T_c$ and domains of ordered phases start to form and grow in time. The equilibrium value of the fluid density $\rho^{eq}$ for the two fluids $\lambda$ and $\bar{\lambda}$ is usually found through simulations; in our case, we obtain $\rho^{eq}_{\lambda,\bar{\lambda}}\simeq 0.1,1.9$. Hence, the order parameter $\phi=(\rho_{\lambda}-\rho_{\bar{\lambda}})/(\rho_{\lambda}+\rho_{\bar{\lambda}})$ ranges approximately between $-0.9$ to $0.9$. 

\section{Results and discussion}

In this section we discuss the numerical results of a phase-separating binary fluid with colloidal dumbbells confined at the interface. As benchmark test, we first study the effect of an isolated dumbbell adsorbed at a flat fluid interface (whose curvature is zero) and afterwards we investigate how a medium/low particle volume fraction affects the curvature of an isolated fluid droplet. We then study the time evolution of the fluid interface during phase separation for different values of particle volume fraction and particle anisotropic ratio and later we extend the investigation to the interface curvature, an observable which, although usually neglected, unveils intriguing properties of the system.

\subsection{Flat fluid interface and isolated droplet}

\begin{figure*}
\centerline{\includegraphics[width=0.9\textwidth]{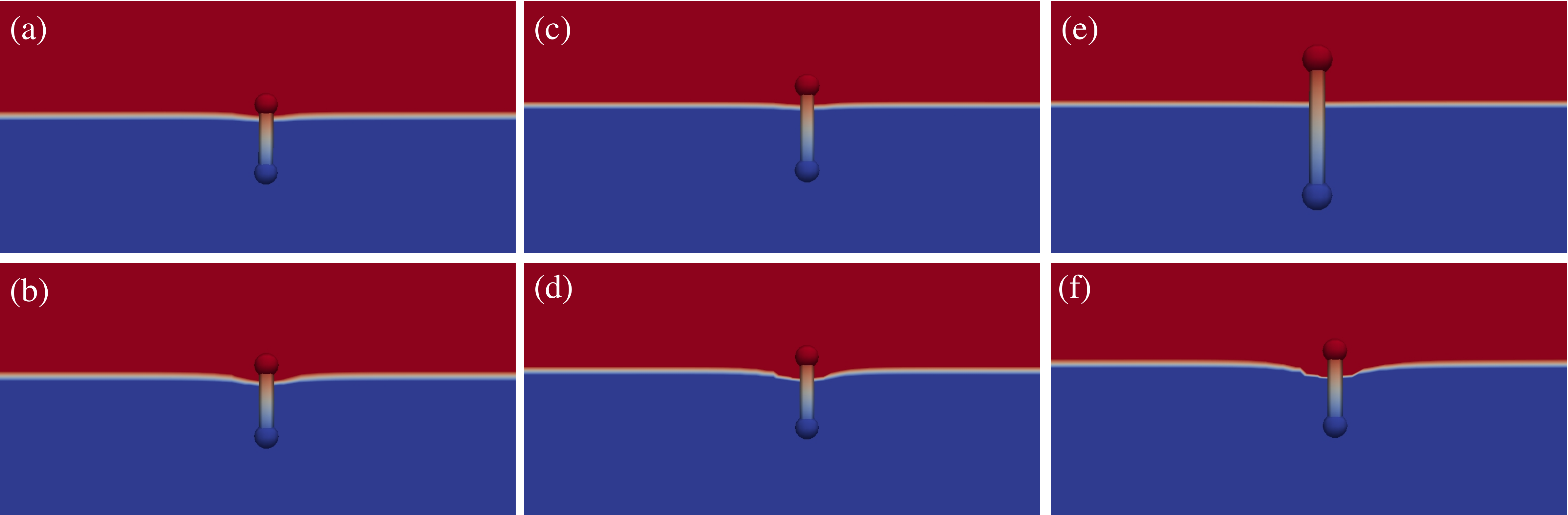}}
\caption{Equilibrium configurations of an isolated dumbbell adsorbed at a flat fluid interface separating two immiscible fluids, the red one (top) and the blue one (bottom) in each figure. Different values of aspect ratio $A$ and solvation coupling constant $\Lambda_k$ are considered.  Top row: $|\Lambda_k|=0.1$ and (a) $A=2$, (c) $ A=2.25$, (e) $A=3$. Bottom row: $A=2$ and (b) $|\Lambda_k|=0.25$, (d) $|\Lambda_k|=0.5$, (f) $|\Lambda_k|=0.75$. For $|\Lambda_k|=0.1$ (top row) the interface slightly bends for $A=2$, and remains almost flat for higher values of $A$. Increasing $\Lambda_k$ (bottom row) favours interface bending, whose deformation is determined by the solvation force,  directed from the red fluid towards the blue one. Simulations are perfomed on a cubic lattice of linear size $L=64$, and each figure represents a 2d perspective of the system obtained by cutting the simulation box at $L_y=32$.}
\label{fig2N}
\end{figure*}

In order to assess how an isolated dumbbell affects the curvature of a flat fluid interface, we consider a phase separated binary fluid and a single dumbbell adsorbed on the resulting interface (see Fig.\ref{fig2N}). If $|\Lambda_k|=0.1$ (top row), the interface is only weakly deformed when $A=2$, whereas for higher values of $A$ it remains approximately flat. This occurs as longer dumbbells have their head and tail placed at larger distance from the fluid interface, which is then less affected by the solvation force. If $|\Lambda_k|$ is higher, the solvation force increases, and the local bend deformation of the interface is more pronounced. The direction of the bend is set by that of the solvation force, which depends on the gradients of the shape function and goes from the red fluid towards the blue one, perpendicularly to the interface. Hence the inclusion of a dumbbell generally increases the interface curvature, and  $\Lambda_k$ can be viewed as a parameter controlling the amount of such increase. Interestingly, while for low values of $|\Lambda_k|$ the interface looks perpendicular to the dumbbell, for higher values its contact angle seems to change. This modification is likely proportional to the ratio $\Lambda_k/g_{\lambda\bar{\lambda}}$, where $g_{\lambda\bar{\lambda}}$ controls the elasticity of the interface. If $\Lambda_k/g_{\lambda\bar{\lambda}}\ll 1$, one would get neutral wetting, whereas with $\Lambda_k/g_{\lambda\bar{\lambda}}\gg 1$  one may achieve complete wetting (or maximal interface deformation).

An approximate mapping between simulation parameters and physical units may be attempted by considering that the typical interface width is $\sim 1\mu$m and occupies $5-7$ lattice sites. Hence one has a lattice step $\Delta x$ correspoding to $\sim 0.2\mu$m, and dumbbells length of $\sim 1\mu$m, comparable with the interface width.

\begin{figure*}
\centerline{\includegraphics[width=0.9\textwidth]{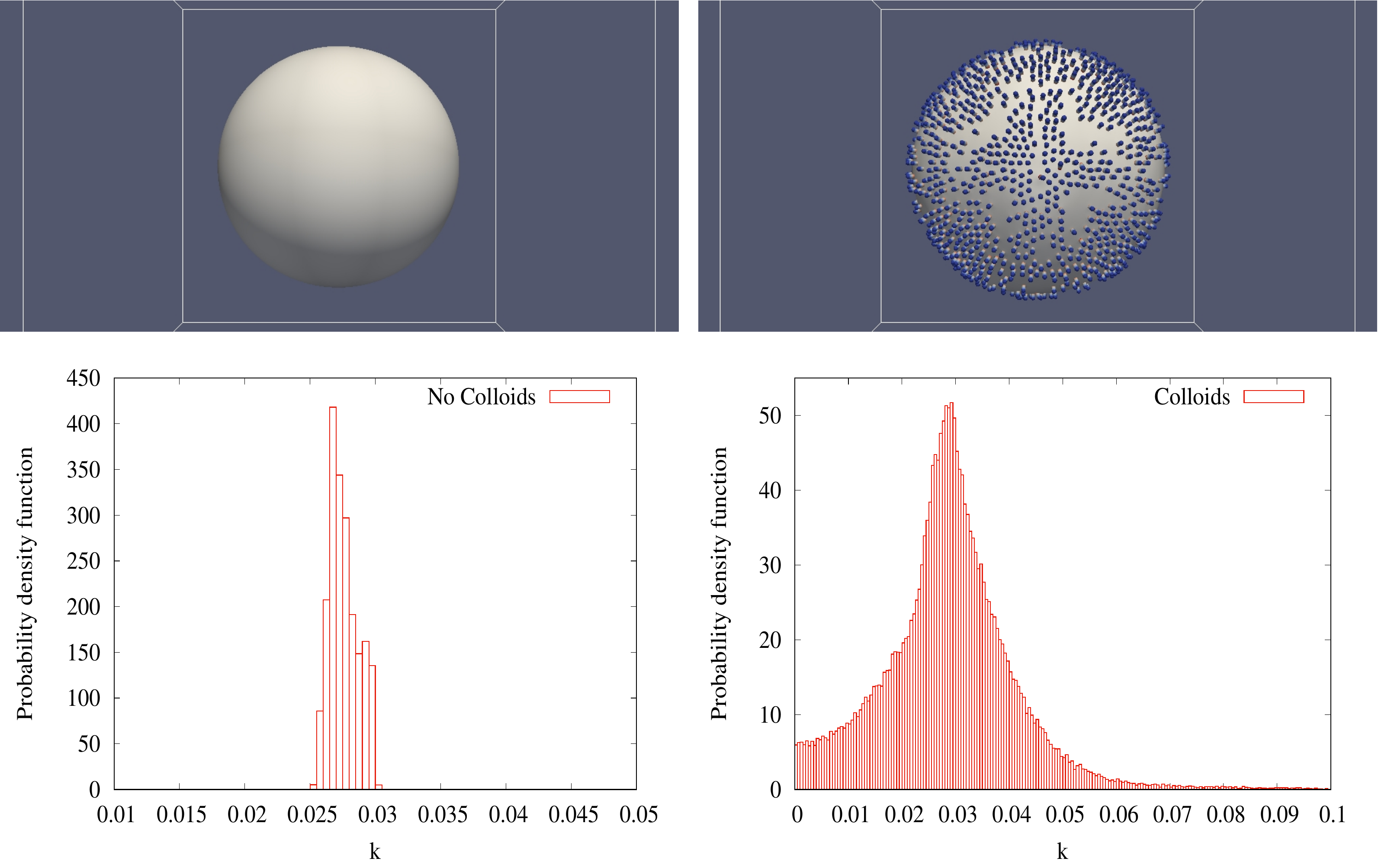}}
\caption{Top left: Isolated fluid droplet at equilibrium ($t=4.5\times 10^{3}\Delta t$) with $V_f=0$. Its geometrical center is located at the center of a cubic simulation box of linear size $L=256$. Top right: Droplet configuration observed when $V_f\simeq 0.006$ (corresponding to $\sim 1500$ dumbbells with $A=2$). Bottom left: Probability density function of the interface curvature observed when $V_f = 0$. A droplet radius of $R \sim 45-50$ lattice sites (including $5-7$ lattice sites of interface width) results. Bottom right: Probability density function of the interface curvature observed when $V_f\simeq 0.006$. Regions of the surface with higher curvature emerge (see the right tail of the distribution), and the droplet attains a spheroidal shape.}
\label{fig3N}
\end{figure*}

A further test proving that the presence of dumbbells generally favours the increase of the fluid interface curvature has been performed by comparing the curvature of a dumbbell-free isolated fluid droplet of radius $R_d$ (whose curvature is constant and equal to $1/R_d$) with that of a droplet on the interface of which a medium/low volume fraction of dumbbells is adsorbed. In Fig.\ref{fig3N} we show two equilibrated configurations of such systems, for the case $V_f=0$ (left) and $V_f\sim 0.006$ (right), respectively. In the latter case dumbells have anisotropic ratio $A=2$ and $|\Lambda_k|=0.1$ (higher values of $\Lambda_k$ may trigger numerical instability). Although slightly higher values of $V_f$ are theoretically feasible, our value represents a reasonable compromise between numerical stability and a high enough packing fraction on the droplet surface.

While with $V_f=0$ the droplet preserves its spherical shape to a very good approximation and the pdf of the interface curvature exhibits a narrow configuration, when $V_f \neq 0$ the droplet acquires a spheroidal shape, with a pdf much broader and a lower peak. The asymmetric shape change of the surface is arguably due to an inhomogeneous distribution of the dumbbells, whose position may change with time due to reciprocal interaction and during droplet equilibration. This test suggests that even a medium/low particle volume fraction can significantly affect the interface curvature and can determine the formation of higher curvature interface regions (see the tail of the pdf) which modify the final droplet shape.

\subsection{Interface dynamics}

\begin{figure*}
\centerline{\includegraphics[width=0.8\textwidth]{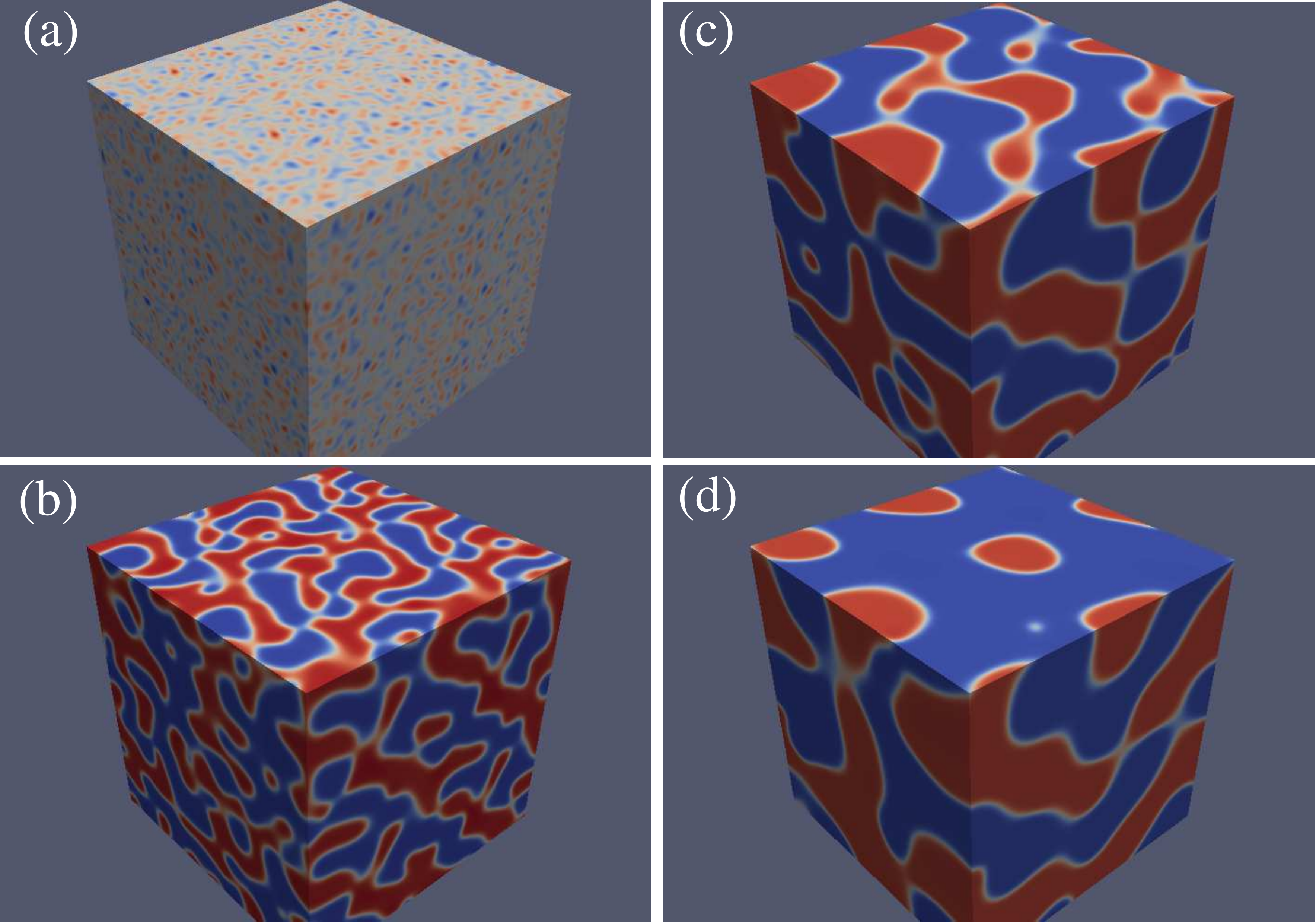}}
\caption{Timelapse of the fluid density of the binary mixture when $V_f=0$, taken at (a) $t=100\Delta t$, (b) $t=1100\Delta t$, (c) $t=3100\Delta t$, (d) $t=5000\Delta t$.} 
\label{fig1}
\end{figure*}

\begin{figure*}
\centerline{\includegraphics[width=0.8\textwidth]{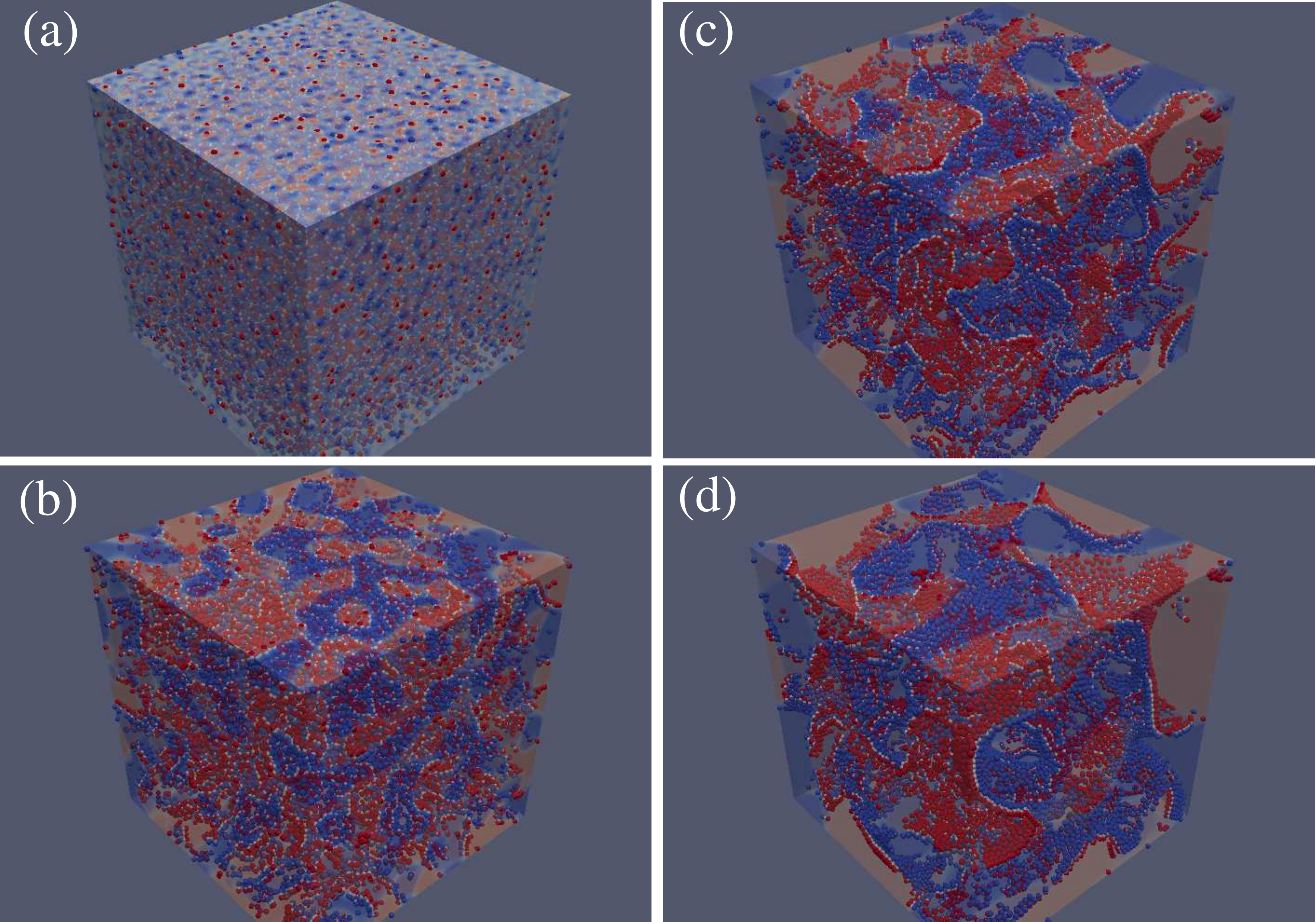}}
\caption{Timelapse of the fluid density of the binary mixture with particle volume fraction $V_f\sim 0.036$. Particles (red and blue spheres linked by a white cylinder), with aspect ratio $A=2.25$ ($d=5$, $r=2$), are anchored at the fluid-fluid (light red and light blue) interface. Snapshots are taken at (a) $t=100\Delta t$, (b) $t=1100\Delta t$, (c) $t=3100\Delta t$, (d) $t=5000\Delta t$.} \label{fig2}
\end{figure*}

We now move on to investigate a more complex system, in which dumbbells are confined at a disordered fluid interface, such as that observed during  the phase separation (more precisely spinodal decomposition) of a binary fluid mixture. We initially consider the phase separation of a dumbbell-free binary fluid (i.e. $V_f=0$, Fig.\ref{fig1}). As expected, domains of each phase, initially of small size (Fig.\ref{fig1}a), grow in time as the phase separation proceeds (Fig.\ref{fig1}b,c), until they attain a sufficiently large size (Fig.\ref{fig1}d), beyond which finite size effects (usually appearing when the characteristic domain size is larger than $1/4$ of the lattice size $L$ \cite{Cates}), become dominant. The inclusion of particles confined at the fluid-fluid interface is expected to significantly affect the dynamics of the fluid. In Fig.\ref{fig2}, we report a simulation of the phase separation of a binary fluid in the presence of dumbbells with $V_f\sim 0.036$ and $A=2.25$. Such value of $A$ sets the longitudinal length of the dumbbell particle at $\sim 10$ lattice points, comparable but slightly larger than the interface width, estimated around $5-7$ lattice spacings. Interestingly, we find that although the dumbbells have weak effects on the coarsening dynamics, they significantly affect the morphology and the curvature of the fluid-fluid interface (see Fig.\ref{fig2}a-d). 

\begin{figure}
\centerline{\includegraphics[width=0.5\textwidth]{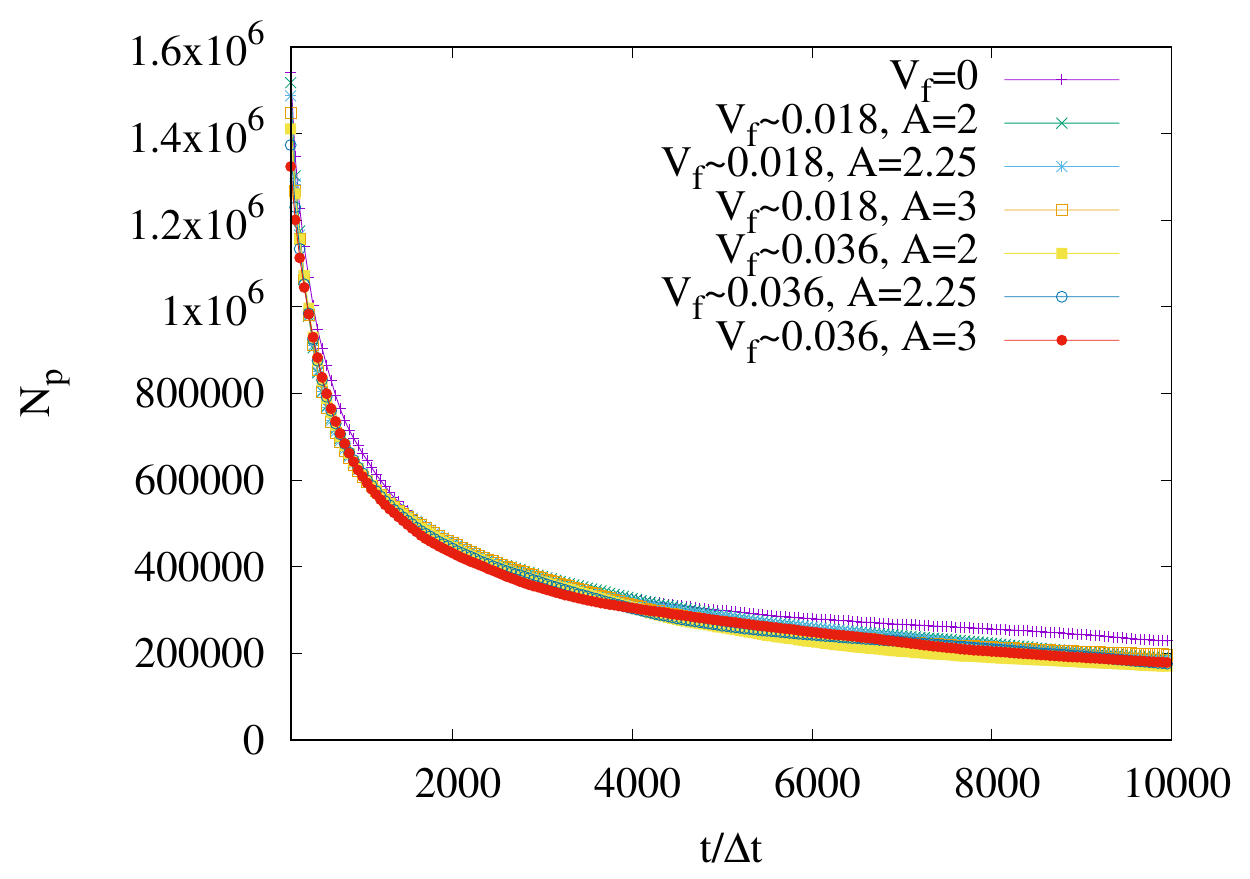}}
\caption{Number of lattice sites $N_p$ where $-0.1<\phi({\bf x},t)<0.1$ $vs$ the simulation time, for three different values of particle volume fraction ($V_f=0$,$ V_f\sim 0.018$, $V_f\sim 0.036$) and aspect ratio ($A=2$, $A=2.25$, $A=3$). Dumbbells slightly favour phase separation ($N_p$ has lower values when $V_f\neq 0$), although the difference with the $V_f=0$ case is rather weak.}
\label{fig3}
\end{figure}

In Fig.\ref{fig3} we provide a more quantitative description of the former behaviour, by inspecting the time evolution of the number of lattice points $N_p$ sitting in the interface region, conventionally defined as $-0.1<\phi({\bf x},t)<0.1$, for different values of dumbbell volume fraction (i.e. $V_f=0$, $V_f\sim 0.018$, $V_f\sim 0.036$) and aspect ratio (i.e. $A=2$, $A=2.25$, $A=3$). We prefer to monitor this quantity rather than other observables, such as the characteristic length of the fluid domain $L(t)$ (see the Supporting Material for the calculation), as it provides direct information on the amount of interface in the system, whose dynamics can be more easily compared with the curvature dynamics (see the next section). While at early times $N_p$ attains high values, as sharp fluid-fluid interfaces occupy large part of the system, at intermediate and late times $N_p$ gradually decreases, due to the progressive loss of interfacial area during the coarsening of the binary fluid. Such process is very mildly affected by the presence of the dumbbells, even at sizeable values of the volume fraction and medium/high aspect ratios, although the inclusion of  anisotropic particles seems to slightly favour phase separation.

Note that values of $V_f$ approximately equal to $0.03$ are usually considered large enough to quench the coarsening and arrest  the mixture in a frozen state (often referred as bijel \cite{Cates2,Harting,Harting2}), consisting of colloidal particles jammed at the fluid-fluid interface. However, our system is different from those discussed in \cite{Cates2} and \cite{Harting} in which the spherical colloids are modelled as rigid spheres \cite{Ladd,Ladd2,Ladd3} with specific surface wetting properties. In our case, the dumbbells consist of two spherical colloidal beads, modelled by means of an immersed boundary method, interacting through an elastic potential, and firmly anchored at the interface via solvation interaction. Our dumbbells are then sequestered at the fluid-fluid interface, yet without contributing to the shrinking of interfacial area, which is driven by diffusive and viscous/inertial coarsening.  This is the reason why our dumbbells do not significantly alter the dynamical fate of the system.  The weak speed-up in the dynamics observed in the presence of the dumbbells is due to the solvation interaction, which favours phase separation by attracting one fluid component and repelling the other one. This phenomenon is slightly more intense for particles with a larger aspect ratio, probably because of the additional effect of the (internal) elastic force, holding the beads together and proportional to the distance between their centers of mass. 

Despite this weak effect on the shrinking of the interface, our simulations show that colloids significantly alter the fluid-fluid interface
curvature, an observable only rarely considered to assess the dynamical properties of binary fluids \cite{Henry}. The next section is dedicated precisely to this point.

\subsection{Curvature dynamics}

We now investigate the dynamics of the magnitude of the fluid-fluid interface curvature, $k=|\nabla\cdot(\frac{\nabla\phi}{||\nabla\phi||})|$, where no distinction between positive and negative sign is made. The signed curvature $k_s$ will be briefly analysed before concluding. 

The aim is to assess whether the interface curvature can capture features of the dynamics of the system, such as inhomogeneities in the local particle volume fraction, which are not easily revealed by inspecting the interface coarsening. To this purpose, we first compute the first and second order moments, $<k>$ and $<k^2>$ respectively, of the probability density function (pdf) of $k$. 

\begin{figure*}
\centerline{\includegraphics[width=1.0\textwidth]{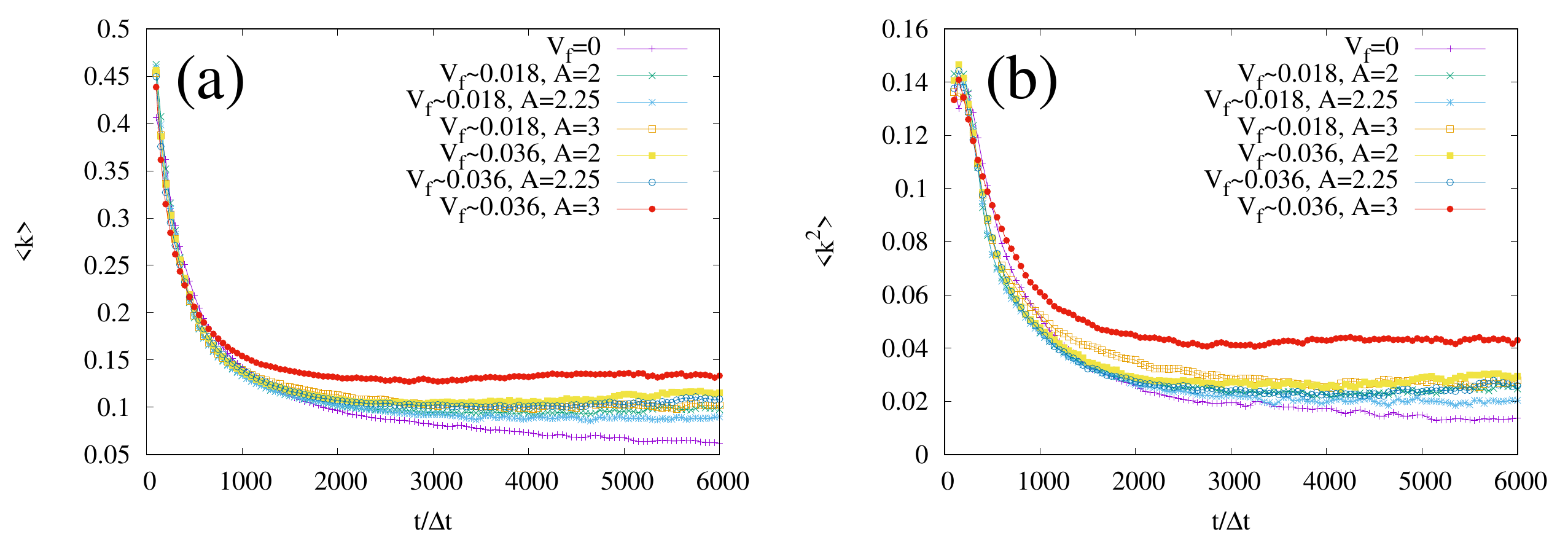}}
\caption{Time evolution of the first moment $<k>$ (a) and of the second moment $<k^2>$ (b) of the fluid-fluid interface curvature, for three different values of particle volume fraction ($V_f=0$,$ V_f\sim 0.018$, $V_f\sim 0.036$) and of particle aspect ratio ($A=2$, $A=2.25$, $A=3$). All functions can be fitted by a stretched exponential time decaying function $e^{-t^{\theta_1}}$, which gives the following values of $\theta_1$: (a) for $V_f=0$ $\theta_1\simeq 0.5$, for $V_f\sim 0.018$ $\theta_1\simeq 0.74$ if $A=2$, $\theta_1\simeq 0.56$ if $A=2.25$ and $\theta_1\simeq 0.6$ if $A=3$ and for $V_f\sim 0.036$  $\theta_1\simeq 0.8$ if $A=2$, $\theta_1\simeq 0.68$ if $A=2.25$ and $\theta_1\simeq 0.67$ if $A=3$. In (b) $\theta_1$ is slightly lower than $1$ for $V_f=0$ and $V_f\sim 0.018$ and slightly higher than $1$ for $V_f\sim 0.036$.}
\label{fig4}
\end{figure*}

In Fig. \ref{fig4}, we show the time evoluton of $<k>$ and $<k^2>$ for different values of particle volume fraction and particle aspect ratio. While at early times high values suggest that the fluid-fluid interface displays pronounced bends throughout the system, later on, both $<k>$ and $<k^2>$ rapidly relax at a rate depending upon $V_f$ and $A$, to finally attain an approximately constant value, generally higher for larger values of $V_f$ and $A$. This is in line with previous studies in which bijels are obtained by including anisotropic colloids (such as rod particles)\cite{Gunther,Gunther2,Clegg}. Such particles are found to decrease the fluid domain size (hence to increase the curvature), as, due to their shape, they occupy a larger interfacial area with respect to spherical colloids. A more detailed analysis of the time evolution of $<k>$ shows that this quantity follows a stretched exponential decaying law $e^{-t^{\theta_1}}$, with $\theta_1\sim 0.5$ if $V_f=0$, with $\theta_1$ increasing from $0.6$ up to $0.8$, going from $A=2$ to $A=3$. For $<k^2>$, $\theta_1$ results slightly lower than $1$ if $V_f=0$ and $V_f \sim 0.018$ and approximately $1.2$ if $V_f\sim 0.036$. Higher value of $<k>$ observed for increasing $V_f$ are tentatively interpreted as the result of the solvation force exerted by each dumbbell adding up to surface tension, thus increasing the effective Laplace pressure, hence the interface curvature. Indeed, the (steady-state) drop across the interface was indeed found to increase, at increasing $V_f$, from about $3.13$ to about $3.54$. This picture may significantly change if adsorption/desorption of particles occurs at the interface. A high adsorption rate may stabilize fluid domains, such as in a bijel\cite{Cates,Clegg}, by diminishing the surface tension, whereas, if particle's desorption dominates, phase separated domains would grow by decreasing interface curvature.

These results support the view that, rather than following the dynamics of $N_p$, in which all curves display an almost identical kinetic pathway (see Fig.\ref{fig3}), it might be more suitable to analyse the time  evolution of the  curvature $k$, since this latter can capture inhomogeneities of the particle concentration, due to local variation of the curvature of a fluid-fluid interface.  This may unveil potential new routes to design functional gradient materials in which heterogeneities are optimised to deliver optimal mechanical performance under a broad variety of load conditions \cite{miyamoto,bhavar}.

From a fundamental standpoint, the immediate question is whether the stretched exponential time decay may hint at a non-trivial form of the equations governing the statistical dynamics of the curvature, namely the time evolution of  the corresponding pdf, $p(k,t)$. In the next section we investigate more carefully this latter point by studying the time evolution of the pdf and its steady-state behavior. 

\subsection{Probability density function of the fluid-fluid interface curvature and fractional dynamics}

\begin{figure*}
\centerline{\includegraphics[width=1.0\textwidth]{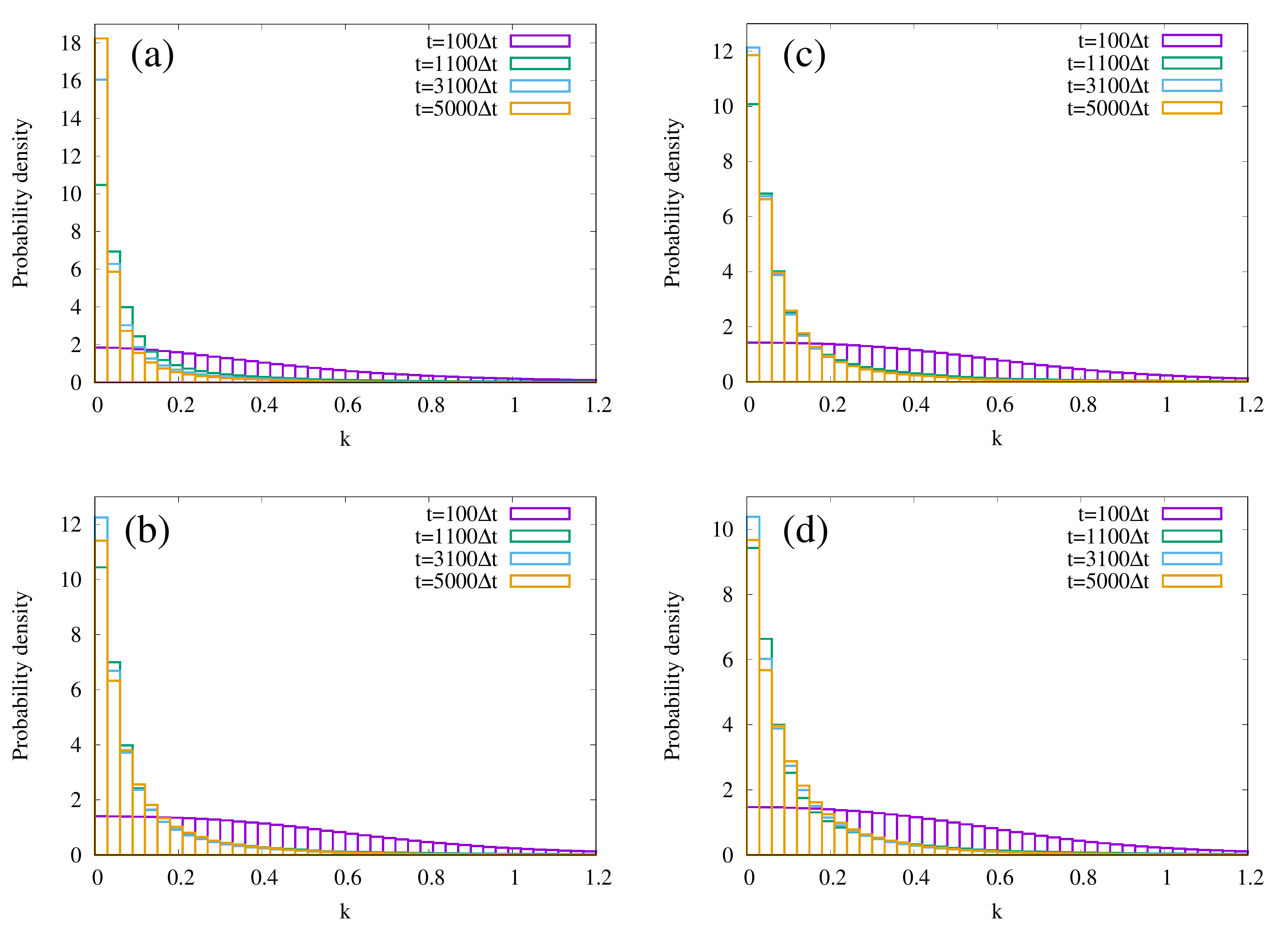}}
\caption{Probability density function of the curvature $k$ for (a) $V_f=0$, (b) $V_f\sim 0.036$, $A=2$, (c) $V_f\sim 0.036$, $A=2.25$, and (d) $V_f\sim 0.036$, $A=3$. Values are taken at $t=100\Delta t$ (purple), $t=1100\Delta t$ (green), $t=3100\Delta t$ (light blue) and $t=5000\Delta t$ (orange).}
\label{fig5}
\end{figure*}

We first consider the time evolution of $p(k,t)$ (the probability density function of $k$ at time $t$) when $V_f=0$ and $V_f\sim 0.036$ (see Fig.\ref{fig5}). At early times ($t=100\Delta t$) $p(k,t)$ appears broadly distributed, as long as $k\leq 0.4$ (as the fluid still contains 
high-curvature stretches), and then gently decays to $0$. Subsequently, for $t\geq 1000\Delta t$, the pdf  collapses towards low values of $k$, indicating that the fluid-fluid interface gets flatter, and rapidly decays to $0$ for high values of $k$. 

The system undergoes a spontaneous smooth transition from an initial disordered state, characterised by broad-shaped pdfs, towards an ordered state described by highly localised distributions. In the limit of a fully flat interface, one would get a Dirac delta $p(k)=\delta(k)$, while for a sphere or radius $R$ one would obtain $p(k) = \delta(k-1/R)$. This is loosely remnant of the "winner-takes-it-all" scenario which characterises a variety of complex systems \cite{Thurner}. Interestingly, such collapse-like dynamics displays a large degree of universality, as it occurs regardless of particle volume  fraction $V_f$ and particle aspect ratio $A$. Eventually, when $V_f=0$, the peak of $p(k)$ is higher as the fluid-fluid interface gets smoother.

\begin{figure}
\centerline{\includegraphics[width=0.5\textwidth]{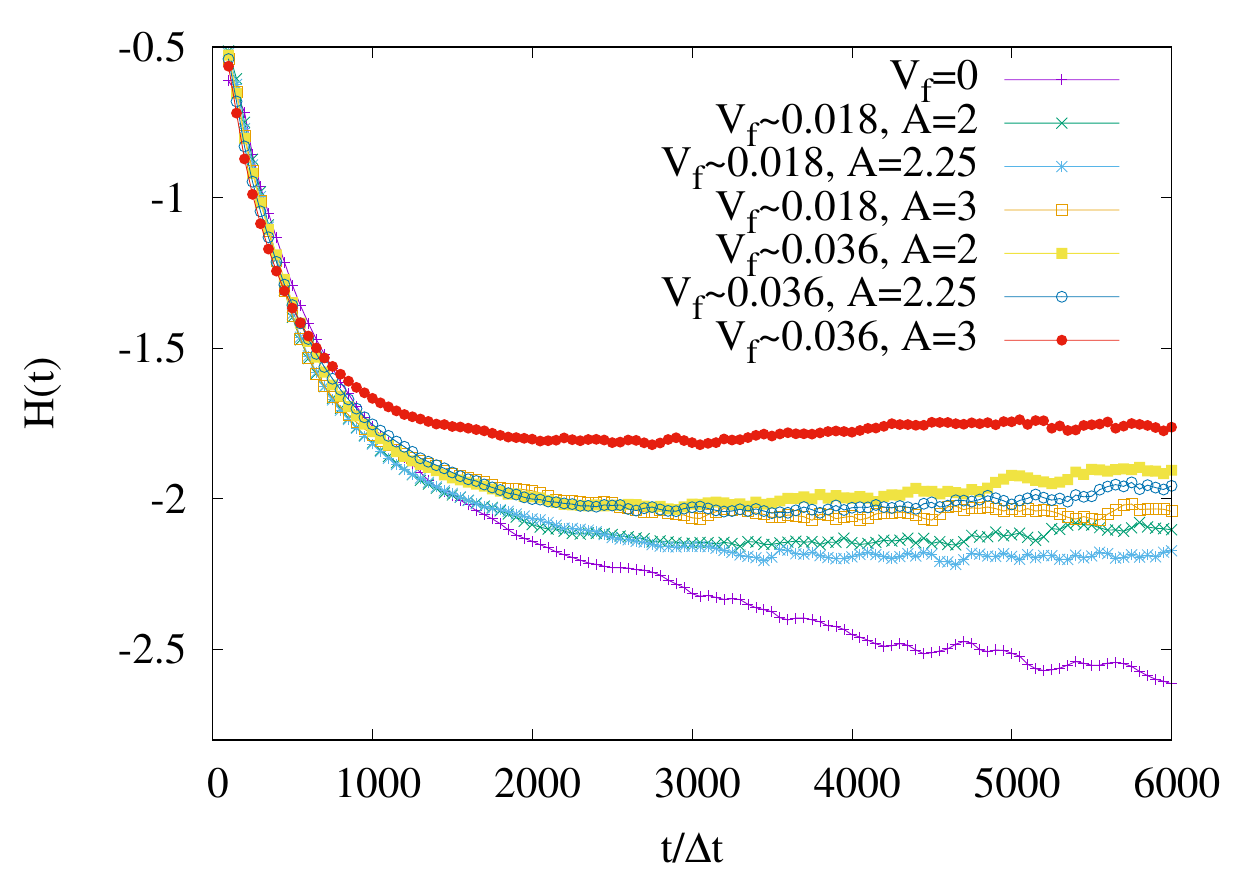}}
\caption{Time evolution of the Kullback-Leibler entropy $H(t)$ for three different values of particle volume fraction ($V_f=0$,$ V_f\sim 0.018$, $V_f\sim 0.036$) and aspect ratio ($A=2$, $A=2.25$, $A=3$). In all cases the entropy is a decreasing monotonic function of time. If $V_f\neq 0$ $H(t)$ remains approximately constant  when $t\geq 3000\Delta t$.}  
\label{fig6}
\end{figure}

The ``distance'' between the late-time and the early-time $p(k,t)$  is best assessed by inspecting the Kullback-Leibler divergence (or relative entropy) $H(t)$ \cite{Kullback}, defined as 
\begin{equation}
H(t)=-\int_0^{k_{max}}  p(k,t)\ln\left(\frac{p(k,t)}{p(k,0)}\right)dk,
\end{equation}
where $p(k,0)$ is the pdf at time $t=0$ and the integral is calculated over all values of $k$, ranging between $0$ and $k_{max}=2$ in our simulations. In all cases, $H(t)$ is a decreasing monotonic function of time (see Fig.\ref{fig6}), as long as $t\simeq 3000\Delta t$ (if $V_f\neq 0$), i.e., as long as the spontaneous transition occurs.  Afterwards, $H(t)$ attains a constant value, once again dependent upon $V_f$ and $A$, indicating that changes of interface curvature become negligible and late-time distributions of $p(k,t)$ do not undergo substantial modifications. 

All results discussed so far suggest that there might be significant scope for monitoring interface curvature as a potentially new probe of the complex physics underlying phase-separating fluids. But what exactly is the interface dynamics? And how is it affected by the inclusion of colloids? Such questions cannot be easily answered experimentally, yet they are key to control the rheological properties of such systems, and the mechanical properties of the materials that can be designed thereof.

To this regard, it is of utmost importance to develop a model of the interface statistical dynamics, namely a kinetic equation for the pdf of the curvature. To this purpose, it is useful to investigate the functional form of $p(k,t)$ at the steady state, approximately attained  at $t\simeq 5000\Delta t$ in our simulations. 

\begin{figure*}
\centerline{\includegraphics[width=1.0\textwidth]{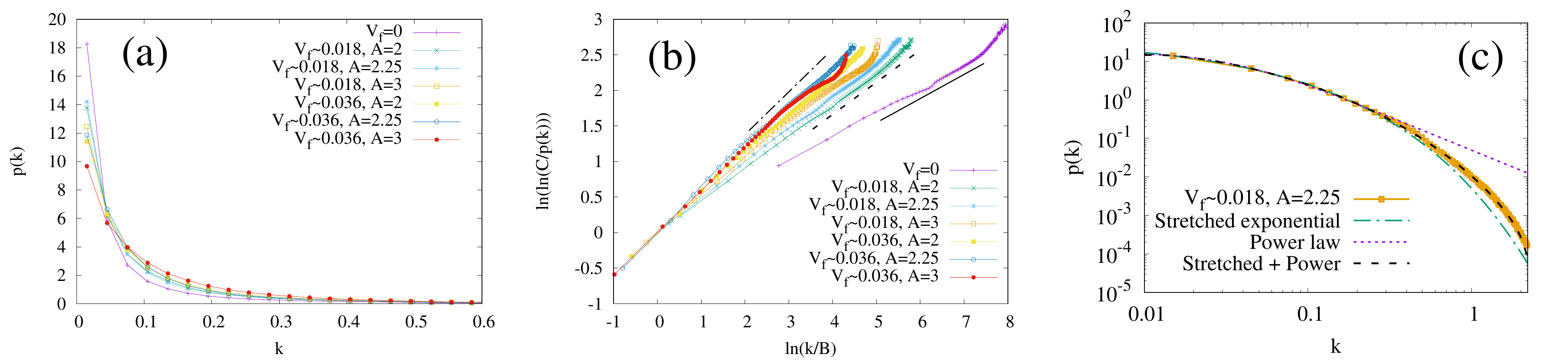}}
\caption{(a) Steady state profile of $p(k)$ for three different values of particle volume fraction ($V_f=0$,$ V_f\sim 0.018$, $V_f\sim 0.036$) and aspect ratio ($A=2$, $A=2.25$, $A=3$) taken at $t=5000\Delta t$. (b) Fit of the steady state profile of $p(k)$ with a stretched exponential function, $f(k)=Ce^{-(k/B)^{\theta_2}}$, where $C$, $B$ and $\theta_2$ are fitting paramters. For $V_f=0$ $\theta_2\simeq 0.34$. For $V_f\sim 0.018$ $\theta_2\simeq 0.45$ if $A=2$, $\theta_2\simeq 0.49$ if $A=2.25$ and $\theta_2\simeq 0.53$ if $A=3$. For $V_f\sim 0.036$ $\theta_2\simeq 0.57$ if $A=2$, $\theta_2\simeq 0.64$ if $A=2.25$ and $\theta_2\simeq 0.6$ if $A=3$. Continuum, dashed and dot-dashed lines are a guide to the eye indicating the slope of purple (plusses), green (crosses) and red (filled circles) curves. (c) Fit of the steady state profile of $p(k)$ for $V_f\sim 0.018$ and $A=2.25$ with a stretched exponential (green dash-dotted line), a power law $k^{-\xi}$ (purple dotted line) and a linear combination of the two, $e^{-(k)^{\psi}}+k^{-\mu}$ (black dashed line), on a log-log scale. One has $\xi\simeq 1.74$ for the power law, and $\psi\simeq 0.38$ and $\mu\simeq 1.86$ for the linear combination. The latter slightly improves the fit of the stretched exponential.}
\label{fig7}
\end{figure*}

In Fig.\ref{fig7}a, we show the late time (and steady state) configuration of $p(k)$ for different values of $V_f$ and $A$. Similarly to what found for $<k>$ and $<k^2>$, we find again that the relaxation dynamics can be described in terms of a stretched exponential function, this time in curvature space, through
\begin{equation}
p(k)=Ce^{-(k/B)^{\theta_2}},
\end{equation}
where $C$, $B$ and $\theta_2$ are fitting parameters. Our simulations show that $\theta_2$ is always significantly lower than $1$, with $\theta_2\simeq 0.34$ for a binary fluid with $V_f=0$, and ranging from $\simeq 0.45$ to $\simeq 0.65$ when $V_f\neq 0$ (see Fig.\ref{fig7}b). 

Since, for low values of $k$ ($k<1$), stretched exponentials are akin to power laws with functional form proportional to $k^{-\xi}$, one may also attempt to fit the steady state profile of the interface curvature with the latter. In Fig.\ref{fig7}c we report a fit of $p(k)$ for $V_f\sim 0.018$ and $A=2$ with both functions and with a linear combination of both. Although the stretched exponential and the power law fit reasonably well $p(k)$ for $k<1$, the former reproduces the decaying bahavior more accurately. Interestingly, an almost perfect fit  stems from a linear combination of both functional forms, and the values of the exponents appearing in each function ($\psi\simeq 0.38$ and $\mu\simeq 1.86$ for the streched exponential and the power law, respectively, see the caption of Fig.\ref{fig7}), are overall comparable with $\theta_2$ and $\xi$, whose values are $0.53$ and $1.74$.

It is known that a power law behavior of the steady state of $p(k)$ could be associated with a fractional Fokker-Planck equation governing its dynamics\cite{Metzler,Zaslavsky}, in which, analogously to a Levy fligth, the propagation of a local perturbation at a given interface location would be characterized by a series of long ``flights'' interrupted by a sequence of trapping events, leading to a super-diffusive dynamics. Although this is not the behavior we observe, a stretched exponential, or a more complex functional form (see Fig.\ref{fig7}c), may support the existence of a super-diffusive dynamics of the interface curvature. This would suggest that the fluid-fluid interface acts as a long-range dynamic correlator for the fluid system, in which a perturbation would be communicated to far-apart regions along the interface much more rapidly than across the bulk fluid; pictorially, it is as if the interface would act as a sort of ``synapsis'' for the complex fluid configuration, i.e. a privileged communication channel. This super-diffusive behavior has been already reported in a wide number of complex systems ranging from bacteria\cite{Klafter,Ariel} and turbulent plasma\cite{Balescu,Checkin} to quantum optics\cite{Schaufler,Schaufler2} and single molecule spectroscopy\cite{Zumofen,Barkai}.

From the mathematical viewpoint, super-diffusive dynamics points to fractional Fokker-Planck equations of the form:
\begin{equation}\label{fract}
 \frac{\partial}{\partial t}p(k,t)={_0}D_t^{1-\theta_2}\left(\frac{\partial}{\partial k}V^{\prime}(k)+D_{\theta_2}\frac{\partial^2}{\partial k^2}\right)p(k,t),
\end{equation}
where $V(k)$ is a potential associated to a drift force and $D_{\theta_2}$ is a generalized diffusion constant. The term ${_0}D_t^{1-\theta_2}\equiv (d/dt(_0D_t^{-\theta_2}))$ is the fractional Riemann-Liouville operator \cite{Riemann}, defined as
\begin{equation}
_0D_t^{-\theta_2}p(k,t)\equiv \frac{1}{\Gamma(\theta_2)}\int_0^t dt^{\prime}\frac{p(k,t^{\prime})}{(t-t^{\prime})^{1-\theta_2}},
\end{equation}
which represents the convolution of $p(k,t)$ with a power-law memory kernel \cite{Metzler}. It has been shown that Eq.(\ref{fract}) stems from a generalized master equation of the type \cite{Metzler2}
\begin{equation}
\frac{\partial}{\partial t}p(k,t)=\int_{-\infty}^{\infty}dk^{\prime}\int_0^t dt^{\prime}K(k,k^{\prime};t-t^{\prime})p(k^{\prime},t^{\prime}), 
 \end{equation}
whose kernel $K$, written in the most general form, introduces time and space correlations \cite{Risken}, a crucial requirement for describing long-range interactions. 

While a detailed analysis of such model lies beyond the scope of the present work, it definitely represents a very interesting subject for future investigations.

\subsection{Signed curvature dynamics}
\begin{figure}
\centerline{\includegraphics[width=0.5\textwidth]{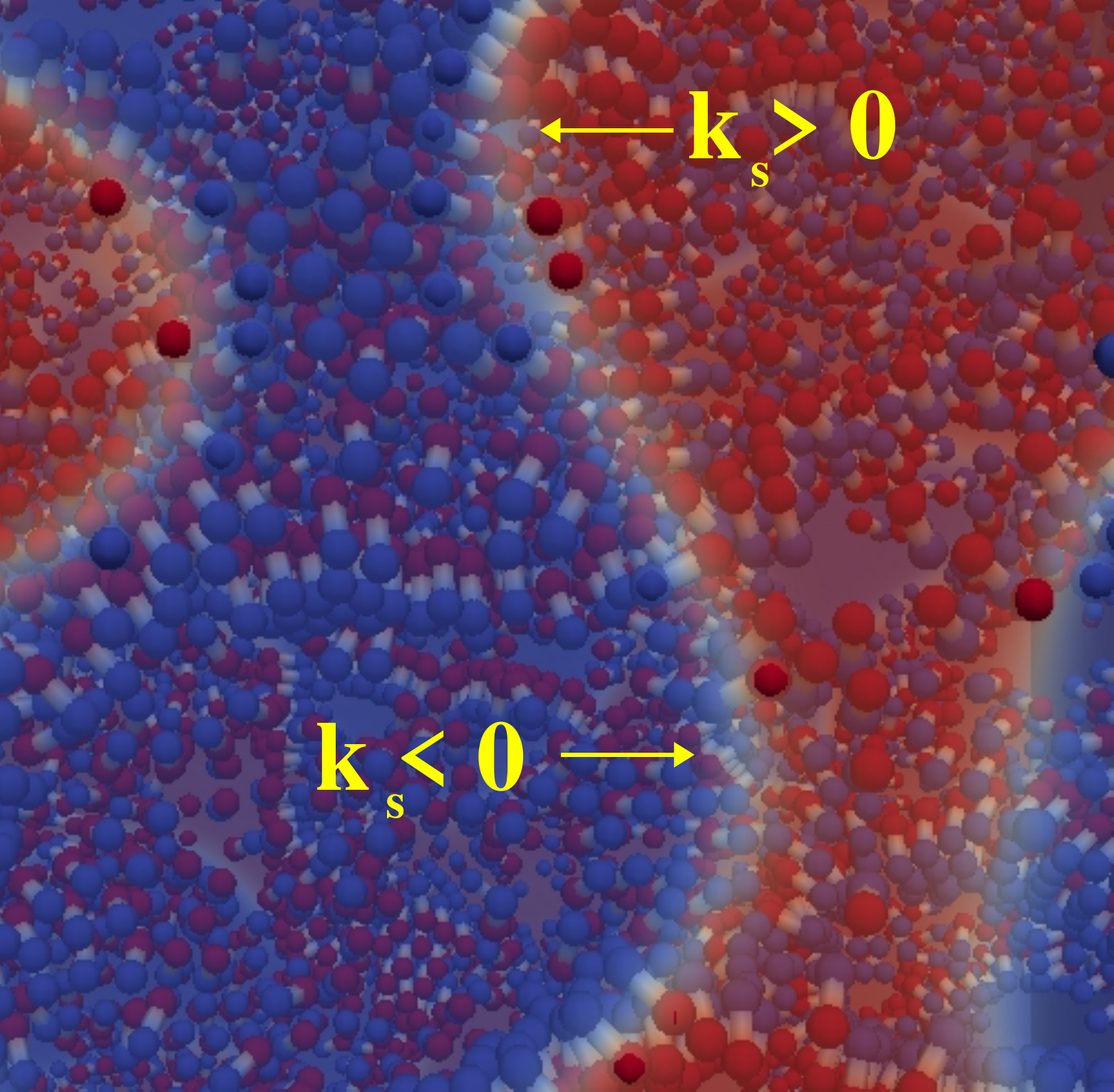}}
\caption{Two-dimensional sketch of the fluid-fluid interface with the corresponding sign of the local curvature. The red component has $\phi\simeq 0.9$, whereas the blue one has $\phi\simeq -0.9$. The region where the interface is approximately flat has $k_s\sim 0$. Colloidal dumbbells are dispersed in the fluid and are anchored at the fluid-fluid interface. Those appearing located within the red and the blue fluid are actually confined at an inner interface, not visible from a 2d perspective.}
\label{fig8}
\end{figure}

In particular, it would be interesting to assess whether, from such mesoscopic dynamics, one can gain insights on the nature of a suitable microscopic model describing the physics of the interface curvature in terms of ``elementary'' interactions between microscopic units of curvature (``curvatons'', for short). Since curvature is a signed quantity, ``curvatons'' would carry themselves a sign, and it is therefore of interest to inspect whether the positive and negative populations display different behaviour in time. In Fig.\ref{fig8}, we show a sketch of the fluid-fluid interface and the corresponding sign of the local curvature.

One may envisage, for instance, a scenario in which localized (and discrete) close-enough regions of interface with opposite sign would merge yielding to an approximately flat interface; such ``curvaton annihilation'' would act as the primary mechanism driving the system towards the minimum surface steady-state configuration.  However, a symmetry-breaking between positive and negative populations must occur whenever the time-asymptotic state results in a positively curved interface, i.e. a spherical droplet.

\begin{figure*}
\centerline{\includegraphics[width=1.0\textwidth]{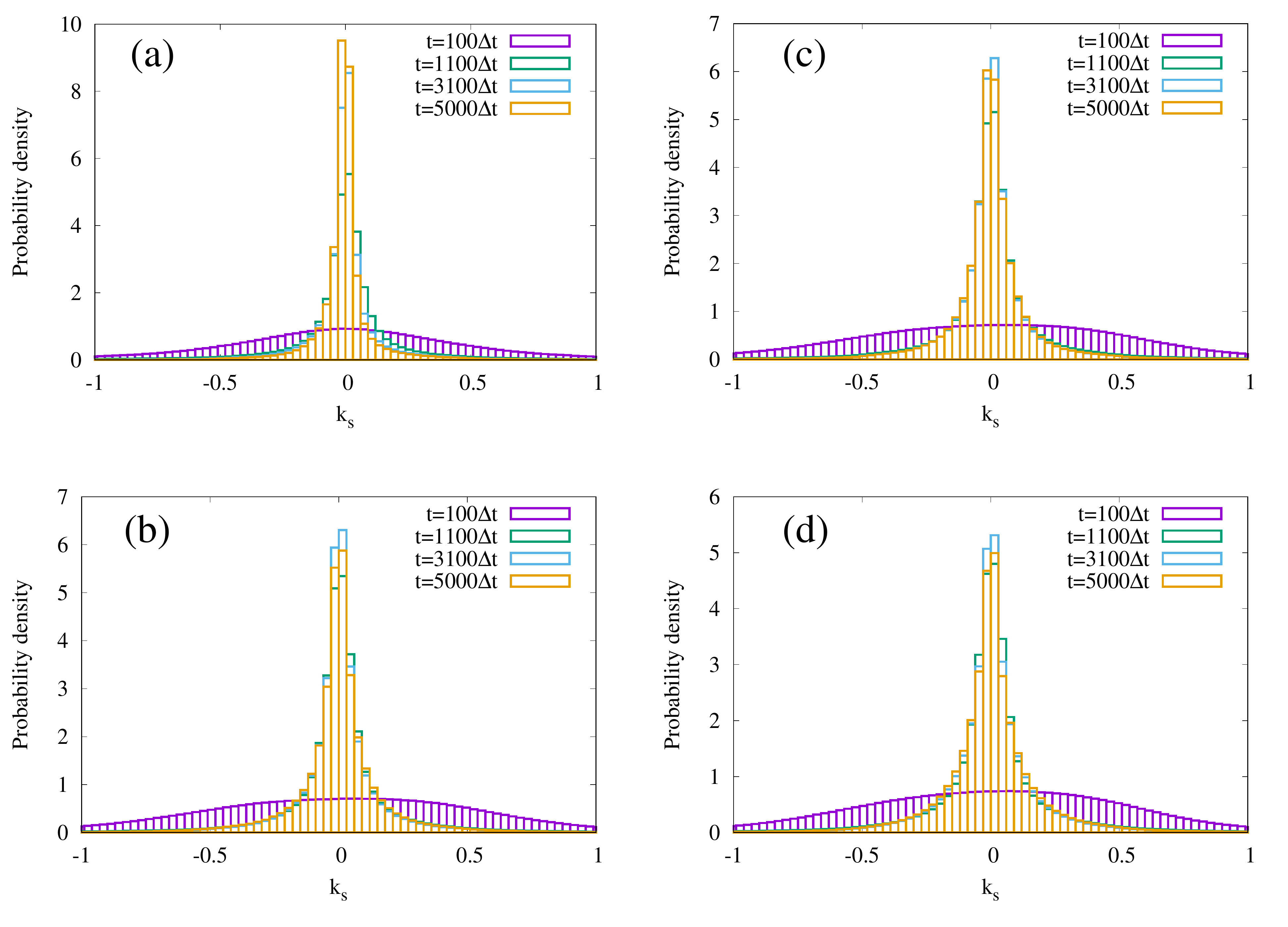}}
\caption{Probability density function of the signed curvature $k_s$ for (a) $V_f=0$, (b) $V_f\sim 0.036$, $A=2$, (c) $V_f\sim 0.036$, $A=2.25$, and (d) $V_f\sim 0.036$, $A=3$. Values are taken at $t=100\Delta t$ (purple), $t=1100\Delta t$ (green), $t=3100\Delta t$ (light blue) and $t=5000\Delta t$ (orange).}\label{fig9}
\end{figure*}

\begin{figure*}
\centerline{\includegraphics[width=1.0\textwidth]{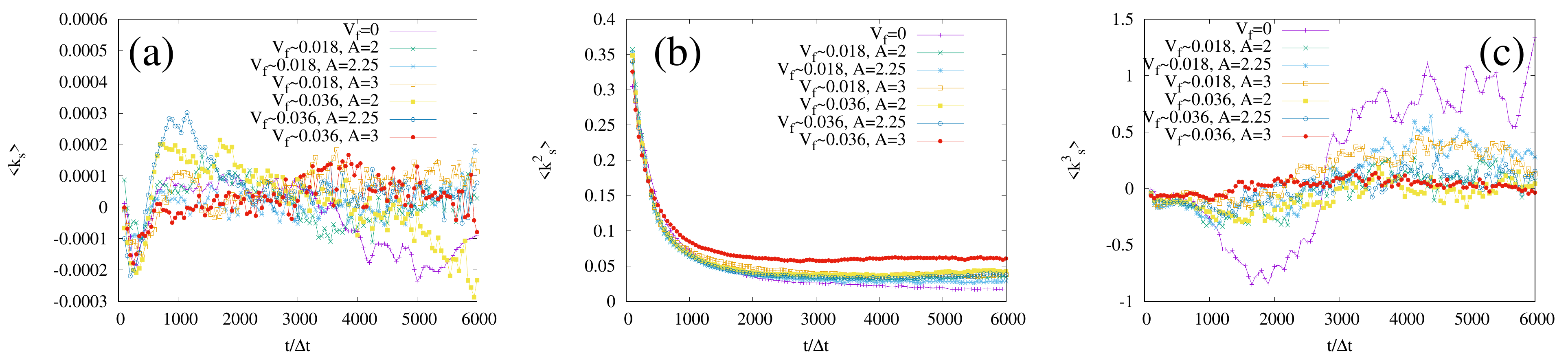}}
\caption{Time evolution of the first order moment $<k_s>$ (a), of the second order moment $<k_s^2>$ (b) and of the third order moment $<k_s^3>$(c) of the signed curvature $k_s$ of the fluid-fluid interface, for three different values of particle volume fraction ($V_f=0$,$ V_f\sim 0.018$, $V_f\sim 0.036$) and of particle aspect ratio ($A=2$, $A=2.25$, $A=3$). Regardless of $V_f$ and $A$, the average roughly fluctuates around zero, and the variance displays a time-relaxation dynamics similar to that observed for the unsigned curvature. The skewness fluctuates around zero for sufficiently high values of $V_f$, while it deviates towards either positive or negative values for weak values of $V_f$.}
\label{fig10}
\end{figure*}

In order to assess whether this picture holds in our case, in Fig.\ref{fig9} we report the time evolution of the signed pdf $p(k_s,t)$. Our results essentially show that there is no compelling evidence of a dominant behavior of one sign (either the positive or negative) of the curvature over the other, at least for high enough values of $V_f$. In other words, interfaces with positive and negative curvature follow a similar dynamics, in which they spontaneously relax from an early stage, more uniform, distribution (i.e. high values of positive and negative curvature almost everywhere in the system) towards a late time unimodal distribution peaked at $k_s\simeq 0$ (i.e. the interface gets flatter). This view is also supported by the time evolution of the average and of the skewness of the distributions, both fluctuating around zero (see Fig.\ref{fig10}). The variance of each pdf, once more, displays a time-relaxing behavior similar to that observed for $k$. 

On the contrary, when $V_f$ is rather low, the pdfs show a non-zero skewness, negative at early times and positive at late times, a sign that negative values of the curvature ``survive'' slightly longer than the positive ones. We attribute this asymmetry to the fact that, while for a sufficiently high values of $V_f$ interface curvature becomes flatter at shorter times (see, for instance, tha variance in Fig.\ref{fig10}b), for low values of $V_f$ the same process takes longer and one sign of the curvature may temporarily dominate over the other. This asymmetry would disappear at very late times or, possibly, by simply running a larger simulation box. We argue that a more robust evidence of asymmetry in the signed curvature could be observed in off-symmetric binary fluid mixtures, in which an emulsion phase would result from the nucleation process \cite{Bray}.

Despite their tentative nature, these considerations indicate that the statistical dynamics of the interface curvature may offer new indicators or order parameters to characterise the long-term behaviour of phase-separating fluids, with and without intersparsed colloids.

\section{Conclusions}

Summarising, we have employed large-scale Lattice Boltzmann simulations to investigate the physics of a fluid-fluid interface in a phase-separating binary fluid and in the presence of colloidal particles. These particles (dumbbells) are modelled by means of two spherical beads, represented via an immersed boundary method, and interacting through an elastic potential. The dumbbells are anchored at the fluid-fluid interface, due to a solvation force, selectively coupling each dumbbell with each phase of the binary fluid. We show that these particles have negligible effects on the interface shrinking, but major ones on the fluid-fluid interface curvature. 

Such effects are significant on the first and on second moment of the probability density function of the curvature, as well as on their steady state dynamics. In particular, we find that the relaxation dynamics of $<k>$ and $<k^2>$ is affected by the particle volume fraction and by their geometry, and their steady state values generally increase at  increasing $V_f$ and $A$. 

A similar effect is revealed by the Kullback-Leibler divergence, a quantity measuring the departure, in terms of relative entropy, between early and late time probability distribution functions.  These latter display a spontaneous and smooth transition from an initial broad-shaped distribution towards  a localised one at late times.

Remarkably, at steady state, the pdfs follow a stretched exponential behavior in curvature space, with an exponent significantly lower than $1$ in all cases. This super-diffusive dynamics sets a strong point towards the picture of the fluid-fluid interface as a self-consistent carrier long--range correlations, what we have symbolically labeled as ``synapses'' of the binary fluid configuration. It is suggested that such long-range correlations may be governed by a fractional Fokker-Planck equation, with a non-local kernel.

Our work sheds light on the mechanism by  which the dynamics the fluid-fluid interface curvature may unveil crucial properties of the systems not easily captured by more popular observables, usually geared towards inspection of the coarse-graining process leading to a minimal-surface long-term configuration. Besides its major theoretical interest, this study is also of potential importance from an experimental standpoint, as it may offer new clues for the design of mesoscale porous materials with novel inhomogeneous mechanical properties, along the line of Functional Gradient materials. One may envisage, for instance, to exploit long-range effects of the interface to remotely control the motion of colloids and  drive them towards targeted structures, a requirement relevant to many biomedical applications, such as drug delivery as well
as microfluidic devices. It is hoped that the above suggestions may be subject to future experimental test and stimulate new technological questions relevant to the design of future functional soft mesoscale materials.

\section*{Appendix: Average fluid domain size}

\begin{figure}
\centerline{\includegraphics[width=0.5\textwidth]{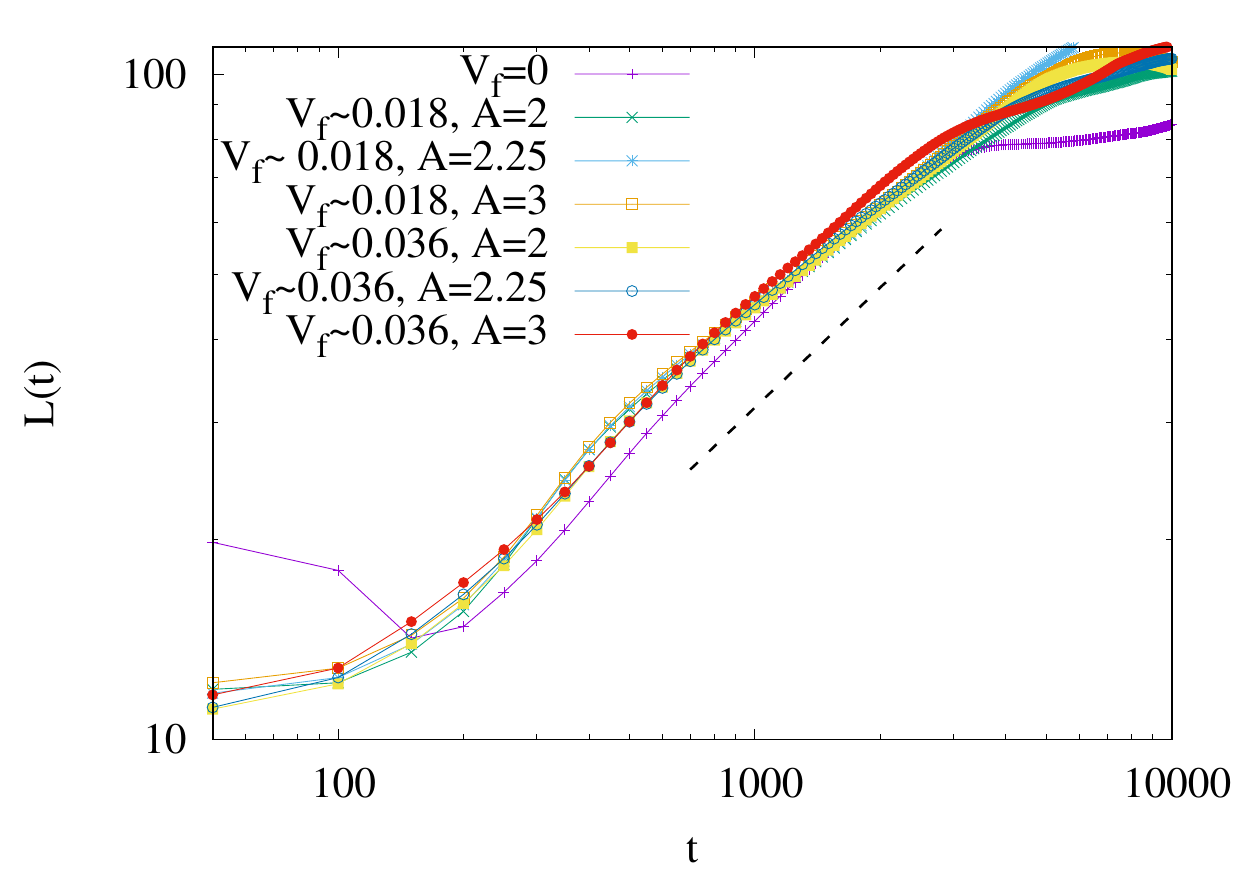}}
\caption{Average fluid domain size $L(t)$ for different values of particle volume fraction $V_f$ and aspect ratio $A$. A slight speed-up of the phase separation dynamics is observed for increasing values of $V_f$ and $A$, with negligible effects on the slope of the curves. Domains grow following a time power law $t^{\Upsilon}$, where  $\Upsilon\simeq 0.6$, within the region from $t\simeq 10^3\Delta t$ to $t\simeq 3\times 10^3\Delta t$. The dashed line, with a slope $0.6$, is a guide to the eye. Log-log scale is set on both axis.}
\label{figS1}
\end{figure}

In this section we plot the average fluid domains size $L(t)$ for different values of particle volume fraction $V_f$ and aspect ratio $A$ (Fig.\ref{figS1}).

By following a standard approach\cite{Cates}, $L(t)$ can be estimated by calculating the inverse of the first moment of the spherically averaged structure factor $S(k,t)=\langle\phi({\bf k},t)\phi(-{\bf k},t)\rangle_k$,
\begin{equation}
L(t)=2\pi\frac{\int S(k,t)dk}{\int kS(k,t)dk},
\end{equation}
where $\phi({\bf k},t)$ is the spatial Fourier transform of $\phi({\bf r},t)$, $k$ is the modulus wave vector of ${\bf k}$ and $\langle\rangle_k$ is an average over a shell in ${\bf k}$ space at fixed $k$.

In all cases domains grow by following a time power law $t^{\Upsilon}$ within a range going from $t\simeq 10^3\Delta t$ to $t\simeq 4\times 10^3\Delta t$. The exponent $\Upsilon$ is found approximately equal to $0.6$, although a slight dynamic speed-up is observed for increasing values of $A$ and $V_f$. We consider the values of $L(t)$ acceptable up to $t\simeq 5.5\times 10^3\Delta t$, after which finite size effects become dominant. Although at $t>3\times 10^3\Delta t$ $L(t)$ is larger than $L/4$, our choice of investigating the physics above such values (but below $t\simeq 5.5\times 10^3\Delta t$) ensures that the interface curvature is reasonably at steady state (see Fig.6-7 of the main text) and finite size effects are acceptably mild.

\section*{Acknowledgements}
The authors acknowledge funding from the European Research Council under the European Union's Horizon 2020 Framework Programme (No. FP/2014-2020) ERC Grant Agreement No.739964 (COPMAT).

\section*{Conflicts of interest} There are no conflicts of interest to declare.






\end{document}